\documentclass[twocolumn, astrosymb]{aastex631} 
\shorttitle{Distinguishing Cusps from Cores using GCs}
\shortauthors{Modak et al.}
\graphicspath{{./}{figures/}}

\usepackage{graphicx}
\usepackage{bm}
\usepackage{hyperref}
\usepackage{float}
\usepackage{amsmath, amssymb}
\usepackage{xcolor}

\newcommand{\Msun}{M_\odot}
\newcommand{\taudf}{\tau_{\mathrm{DF}}}

\newcommand{\dv}[2]{\frac{\mathrm{d}#1}{\mathrm{d}#2}}

\begin{document}

\title{Distinguishing Dark Matter Cusps from Cores using Globular Clusters}

\correspondingauthor{Shaunak Modak}
\email{shaunakmodak@princeton.edu}

\author[0000-0002-8532-827X]{Shaunak Modak}
\affiliation{Department of Astrophysical Sciences, 4 Ivy Lane, Princeton University, Princeton, NJ 08544, USA}

\author[0000-0002-1841-2252]{Shany Danieli}
\altaffiliation{NASA Hubble Fellow}
\affiliation{Department of Astrophysical Sciences, 4 Ivy Lane, Princeton University, Princeton, NJ 08544, USA}

\author[0000-0002-5612-3427]{Jenny E. Greene}
\affiliation{Department of Astrophysical Sciences, 4 Ivy Lane, Princeton University, Princeton, NJ 08544, USA}

\begin{abstract}
Globular Clusters (GCs) provide valuable insight into the properties of their host galaxies' dark matter halos. Using N-body simulations incorporating semianalytic dynamical friction and GC-GC merger prescriptions, we study the evolution of GC radial distributions and mass functions in cuspy and cored dark matter halos. Modeling the dynamics of the GC-rich system in the dwarf galaxy UGC\,7369, we find that friction-induced inspiral and subsequent mergers of massive GCs can naturally and robustly explain the mass segregation of the GCs and the existence of a nuclear star cluster (NSC). However, the multiple mergers required to form the NSC only take place when the dark matter halo is cuspy. In a cored halo, stalling of the dynamical friction within the core halts the inspiral of the GCs, and so the GC merger rate falls significantly, precluding the formation of an NSC. We therefore argue that the presence of an NSC requires a cusp in UGC\,7369. More generally, we propose that the presence of an NSC and the corresponding alteration of the GC mass function due to mergers may be used as an indicator of a cuspy halo for galaxies in which we expect NSC formation to be merger-dominated. These observables represent a simple, powerful complement to other inner halo density profile constraint techniques, and should allow for straightforward extension to larger samples.
\end{abstract}


\section{Introduction}
\label{sec:intro}

The diversity of observed dark matter halo density profiles is one of the key small-scale challenges to $\Lambda$CDM cosmology \citep{LambdaCDM_Review}. While dark-matter-only hydrodynamical $\Lambda$CDM simulations universally produce dark matter halos well-fitted by inner density cusps $\mathrm{d}\log\rho/\mathrm{d}\log r < 0$ \citep{NFW}, observations indicate that some galaxies’ halos have inner cores, where the density approaches a constant value near the center instead of diverging \citep{McGaugh_Rotation_Curves, LITTLETHINGS, Unexpected_Diversity, SantosSantos_Diversity}.

A wide variety of explanations for this ``core-cusp'' discrepancy have been proposed, including self-interacting dark matter models \citep{SIDM}. One possible resolution within the $\Lambda$CDM paradigm involves depletion of mass from the inner regions via baryonic feedback mechanisms not present in CDM-only simulations. The diversity of observations is then explained by the varying efficiency of feedback in different environments and at different stellar masses \citep{Pontzen_Governato2012, CoreCuspReview, Unexpected_Diversity}.

Dwarf galaxies provide excellent laboratories to test these hypotheses, due to their tendency to be dark matter dominated but low enough in mass to be sensitive to stellar feedback. To that end, a sample of dwarf galaxies with well-characterized inner density profile slopes in varied environments is vital. Unfortunately, distinguishing between cores and cusps with stellar kinematics is challenging due to the mass-anisotropy degeneracy (e.g. \citealt{BT, MAD_dwarfs}); current studies primarily use HI rotation curves \citep{LITTLETHINGS, SPARC} and so are restricted in applicability to gas-rich galaxies. In this paper, we consider alternative methods of characterizing dark matter halos using more easily accessible photometric data in the form of GC populations.

Dynamical studies of GC populations have been used to characterize dark matter halos observationally in the past (e.g. \citealt{Tremaine_Nuclei, Tremaine_Fornax}), and recent works have included both entirely semianalytic \citep{Gnedin_NSC, SS_semianalytic} and more realistic N-body simulations of varying degrees of complexity \citep{ColeDehnenReadWilkinson2012, ArcaSedda_Lack_of_NSCs_BHs, Nusser2018, Chowdhury_GC_Baryons, Chowdhury_GC_Nbody, UDG_DF_MOND, UDG1_DF}. The key principle behind all of these analyses is the same: dynamical friction \citep{Chandrasekhar_DF} due to the halo acts on the GCs to cause their orbits to inspiral at a rate dependent on the local halo properties as well as the GC mass \citep{DF_in_Dwarfs}. Therefore, observations of the mass function and radial distributions of the GCs yield constraints on the halo’s density profile. The Fornax dwarf galaxy is one well-studied example: the classical ``GC Timing Problem’’ argues that the dark matter halo of this galaxy must be cored, due to the lack of an NSC. In the presence of a cusp, GC inspiral timescales would be substantially less than a Hubble time, necessarily depositing GCs at the galaxy's center \citep{Tremaine_Fornax, Fornax_MOND, Fornax_Meadows, Fornax_DDM}. \cite{Chowdhury_GC_Baryons} used similar reasoning to argue against the presence of a low-mass cuspy halo in the ultra diffuse galaxy NGC\,1052-DF2. Most recently, \cite{UDG1_DF} conducted simulations demonstrating that the observed ``GC mass segregation’’ (more massive GCs tending to have smaller projected radii) in NGC\,5846-UDG1 could be entirely explained by the increased friction force experienced by more massive clusters. The radial distribution of the GCs could then be used to simultaneously estimate the initial GC formation radius and demonstrate a preference for a massive dark matter halo.

GC mergers are considered to be a critical channel for NSC formation \citep{Tremaine_Nuclei, Neumayer_NSC}. Dynamical friction has been identified as the chief driving force behind these mergers: as the typical orbital radii of the GCs become smaller compared to their sizes, mergers become significantly more likely, so they occur predominantly in galactic nuclei \citep{CapuzzoDolcetta_NSCmergers, CapuzzoDolcetta_NSCsims, Bekki_DFNSCs, Agarwal_NSCs}. Recent observational studies of NSC stellar populations demonstrate a preference for this formation channel over in-situ growth for dwarf galaxies with $M_\star\lesssim 10^9\Msun$ across a wide range of environments \citep{Neumayer_NSC, NSC_formation_dwarfs}.

Modeling the GC population of a nucleated galaxy provides a natural opportunity to investigate both of these phenomena jointly, i.e. to test whether dynamical friction is able to facilitate the inspiral of an appropriate number of individual GCs within the lifetime of the galaxy to produce a massive NSC in a realistic setting. To that end, motivated by the proof-of-concept in \cite{UDG1_DF}, we develop simple N-body simulations implemented in the REBOUND code \citep{rebound}, incorporating models of GC mass loss and GC-GC mergers as well as semianalytic dynamical friction prescriptions that include important effects such as core stalling \citep{Read_core_stalling, Gualandris_stalling, Petts2015_DF}. We aim to use the simulations to constrain the nature of the halo by mimicking observed GC radial distributions and GC mass functions.

Here we present the results of our simulations of UGC\,7369, a nearby early-type dwarf galaxy. Unlike Fornax, UGC\,7369 is nucleated, and because it is relatively GC rich, we are able to gain a clearer understanding of its GC population and thus make more robust constraints on the nature of its halo. Briefly, we argue that the mass and projected orbital radius of the NSC in UGC\,7369 are plausibly explained by GC-merger-dominated formation, but only in a cuspy halo. In contrast to the Fornax timing argument in which the lack of an NSC was interpreted as a sign of a cored halo, here we suggest that the presence of an NSC may be interpreted as a sign of a cusp. We explore how this correlation may be extended to a wide class of other dwarf galaxies as well, and propose both the presence of an NSC and properties of the GC mass function as simple observables that complement kinematic techniques.

In \S\ref{sec:data}, we describe the data used in this work. In \S\ref{sec:sim_design}, we outline the physics incorporated into our simulations and motivate parameter choices for our dynamical models. In \S\ref{sec:results}, we describe how the simulations are compared with observations and present the results of our analysis. Finally, we detail the implications of our results for the core-cusp problem as well as possible generalizations of our work in \S\ref{sec:discussion}, and summarize our main findings in \S\ref{sec:summary}.

\section{Data}
\label{sec:data}

\subsection{Our GC Sample}
\label{sec:sample}

We use the state-of-the-art sample of GC populations associated with nearby ($d = 3 - 12$\,Mpc) dwarf galaxies presented in \citet{Georgiev_sample1, Georgiev_sample3}\footnote{All the HST data used in this paper can be found in MAST: \dataset[10.17909/9fbz-m809]{http://dx.doi.org/10.17909/9fbz-m809}.}. The sample is drawn from archival Hubble Space Telescope (HST) ACS/WFC F606W and F814W images, where the spatial resolution afforded by the ACS allows for a clean photometric identification of the GCs. A typical GC with half-light radius $r_{h}=3\,\mathrm{pc}$ is resolved at the sample galaxies' mean distance of $5\,\mathrm{Mpc}$. Out of the full sample of 68 dwarfs, we restrict our consideration to galaxies with measured distances, measured stellar masses, and at least 1 old GC. Our subset consists of 36 dwarf galaxies with varied morphologies and environments, and stellar masses ranging from $M_\star\sim 7\times 10^5\Msun$ to $2\times 10^9\Msun$.

These galaxies host 161 GCs in total, each of which has measured $V$-band absolute magnitudes, $V-I$ color, and projected separation from the center of its host. \citet{Georgiev_sample1} selected the GCs as objects with colors satisfying $0.7 < V-I < 1.4$, $M_V < -2.5$, and ellipticities $\epsilon < 0.4$. \citet{Georgiev_sample3} estimated the contamination upper limit of the sample to be $0.13$ objects/arcmin$^2$, corresponding to at most $1.5$ objects per ACS field, which are expected to be near the absolute magnitude cut limit.

\begin{figure}
    \centering
    \includegraphics[width=0.47\textwidth]{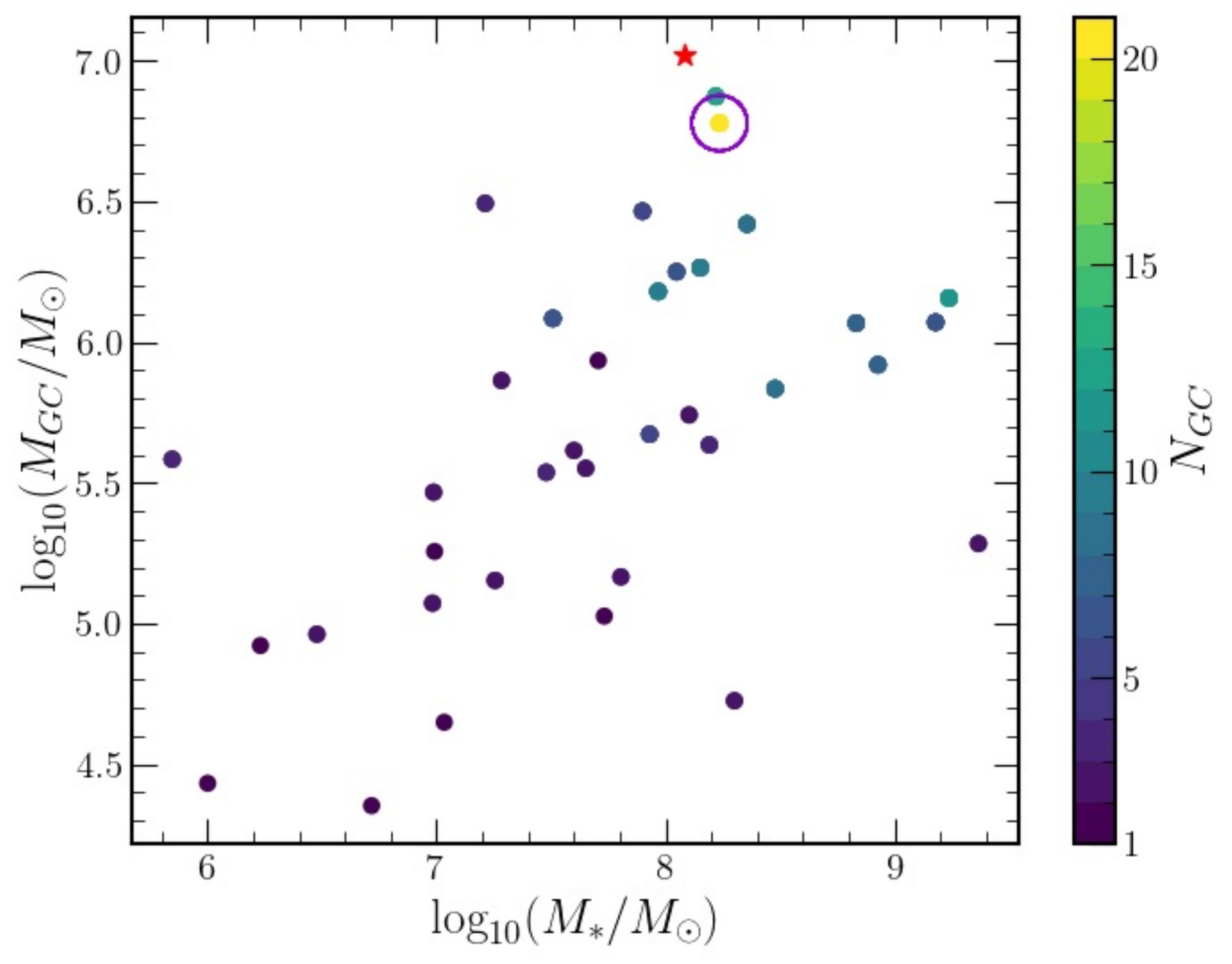}
    \caption{The stellar masses and total mass in GCs of the galaxies in the \citet{Georgiev_sample3} sample, color-coded by the number of GCs present in the system. UGC\,7369, circled in purple, is visible as a clear outlier compared to the remainder of the sample in $N_{GC}$. For comparison, the red star shows the position of NGC\,5846-UDG1 (studied in \citealt{UDG1_GCs, UDG1_DF}), which has an even higher $N_{GC}=33$, in this plane.}
    \label{fig:MGC_Mstar}
\end{figure}

A central input we need to model the GC population is its mass function. We therefore compute the stacked GC mass function for the GCs associated with our sample of 36 dwarfs. First, we convert the absolute magnitudes to masses using a color-dependent mass to light ratio ($M/L_V$) derived using the relation from \citet{Roediger_Courteau_color_ML}. Because these relations are calibrated on galaxies, we expect to slightly overestimate $M/L_V$ at a fixed color, leading to a systematic increase in measured GC masses \citep{Villaume_2017}, but we do not expect such a shift to substantially alter our conclusions. For reference, the absolute magnitude and color cuts involved in sample selection correspond to a minimum GC mass of $\sim 500\Msun$, while the minimum mass of a GC in the final sample is $\sim 10^4\Msun$. Figure \ref{fig:MGC_Mstar} summarizes the sample of galaxies in the total GC mass ($M_{GC}$) -- host galaxy stellar mass ($M_\star$) plane. Across the sample, the average $N_{GC} = 4.3$ clusters, and the average $M_{GC} = 9.4\times 10^5\Msun$.

\begin{figure}
    \centering
    \includegraphics[width=0.47\textwidth]{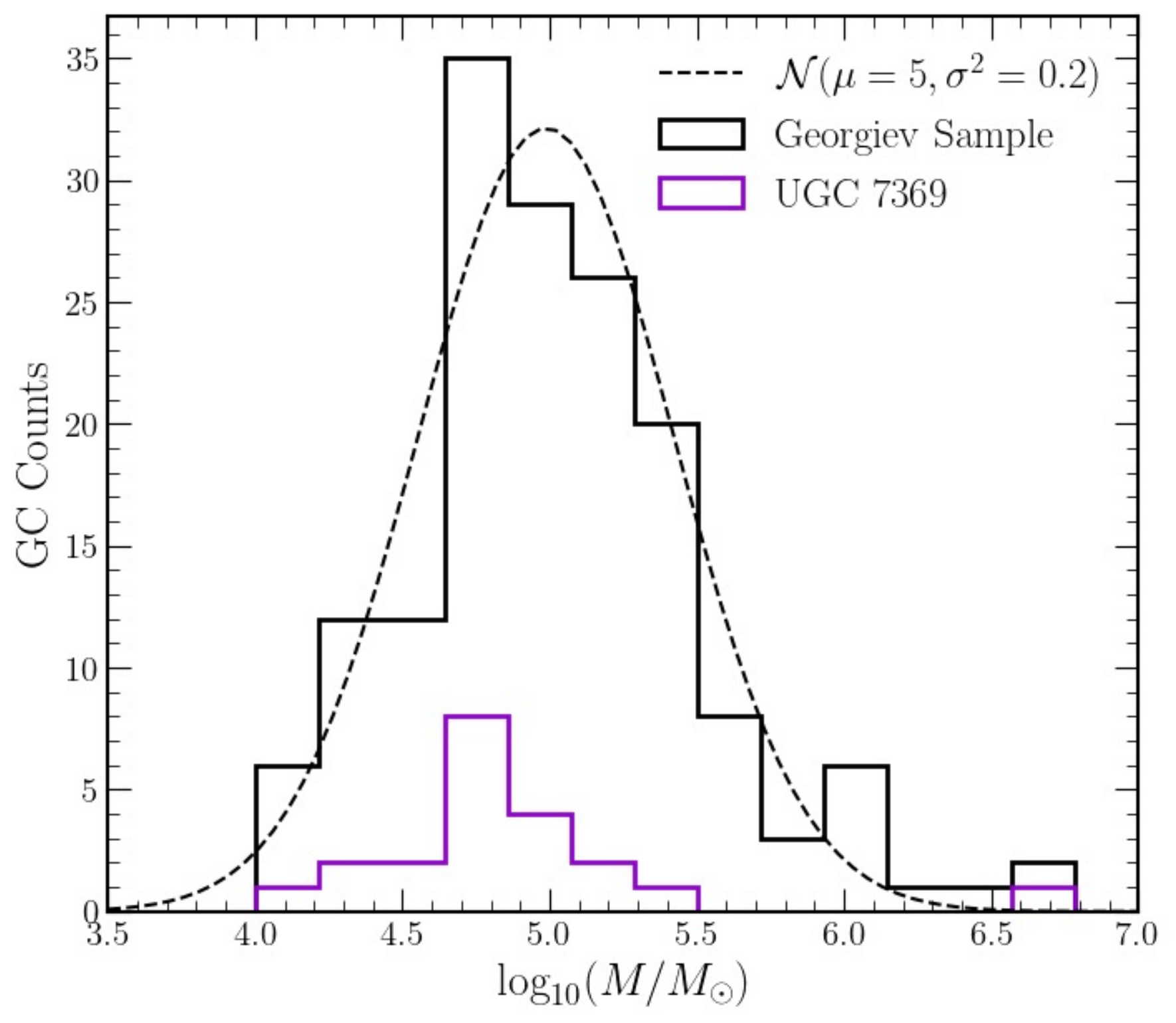}
    \caption{The GC mass function from the \cite{Georgiev_sample1} catalogue and a log-normal fit to the data that excludes the NSCs in the most-massive bin. The mass function of the GC population in UGC7369 is shown in purple for comparison.}
    \label{fig:GCMF}
\end{figure}

The distribution of GC masses across all 36 sample galaxies is displayed in Figure \ref{fig:GCMF}. Over a wide range of stellar masses and distances, the GC luminosity function is observed to take a ``universal'' form, and a common form both theoretically-motivated and observed for this distribution is log-normal (see e.g. \citealt{Hanes_GCLF, Brodie_Strader_GC_Review, Rejkuba_GCLF, Harris_GCLF}). Although we do not use a uniform mass-to-light ratio, we still find that the sample's mass function is well-fitted by a log-normal distribution, centered at $M = 10^5\Msun$ with a variance of $\sigma^2 = 0.2$. In our simulations, presented in \S\ref{sec:sim_design}, we assume that GC masses may initially be treated as independent random draws from this distribution, and treat observed deviations from this distribution as a property of the system that must be produced dynamically as the system evolves. Although other forms for the universal mass function have been proposed, we find that the log-normal distribution used here provides a reasonable fit, and we expect that more complex functional forms would not alter our conclusions appreciably, as long as they are not substantially more top- or bottom-heavy (see \S\ref{sec:degeneracies} for a more detailed discussion).

Overall, our final dataset includes the normalized projected radii $r_\perp/R_\mathrm{e}$ of each GC as well as its mass $M$. Although we use the entire sample to motivate simulation parameter choices and contextualize our results, we focus our analysis on one particularly GC-rich galaxy, UGC\,7369.

\subsection{UGC\,7369}
\label{sec:UGC7369}

UGC\,7369 is a lenticular dwarf galaxy with a measured stellar mass of $M_\star = 1.72\times 10^8 \Msun$ \citep{Georgiev_sample3}. As demonstrated in Figure \ref{fig:MGC_Mstar}, this galaxy is notable for its GC richness. It hosts 21 GCs, by far the most out of any galaxy in this sample, which is important because segregation effects are only robustly detectable in a sufficiently large population of varied GC masses. Across its 21 clusters, it includes a total GC mass of $M_{GC} = 6\times 10^6\Msun$, yielding an above-average $M_{GC}/M_\star$ ratio of 0.035. Moreover, its identified GCs all satisfy $M_V < -4.95$ mag, so being on the brighter end of the sample, they are less susceptible to contamination; according to \cite{Georgiev_sample3}, contamination from background galaxies accounts for at most one of the GCs. Compared to the rest of the sample, UGC\,7369's higher total number of GCs (and hence a higher total mass in GCs) also represents a more thorough sampling of the universal mass function, allowing for better probing of how the mass function evolves to arrive at the present day distribution shown in purple in Figure \ref{fig:GCMF}.
\begin{figure}
    \centering
    \includegraphics[width=0.47\textwidth]{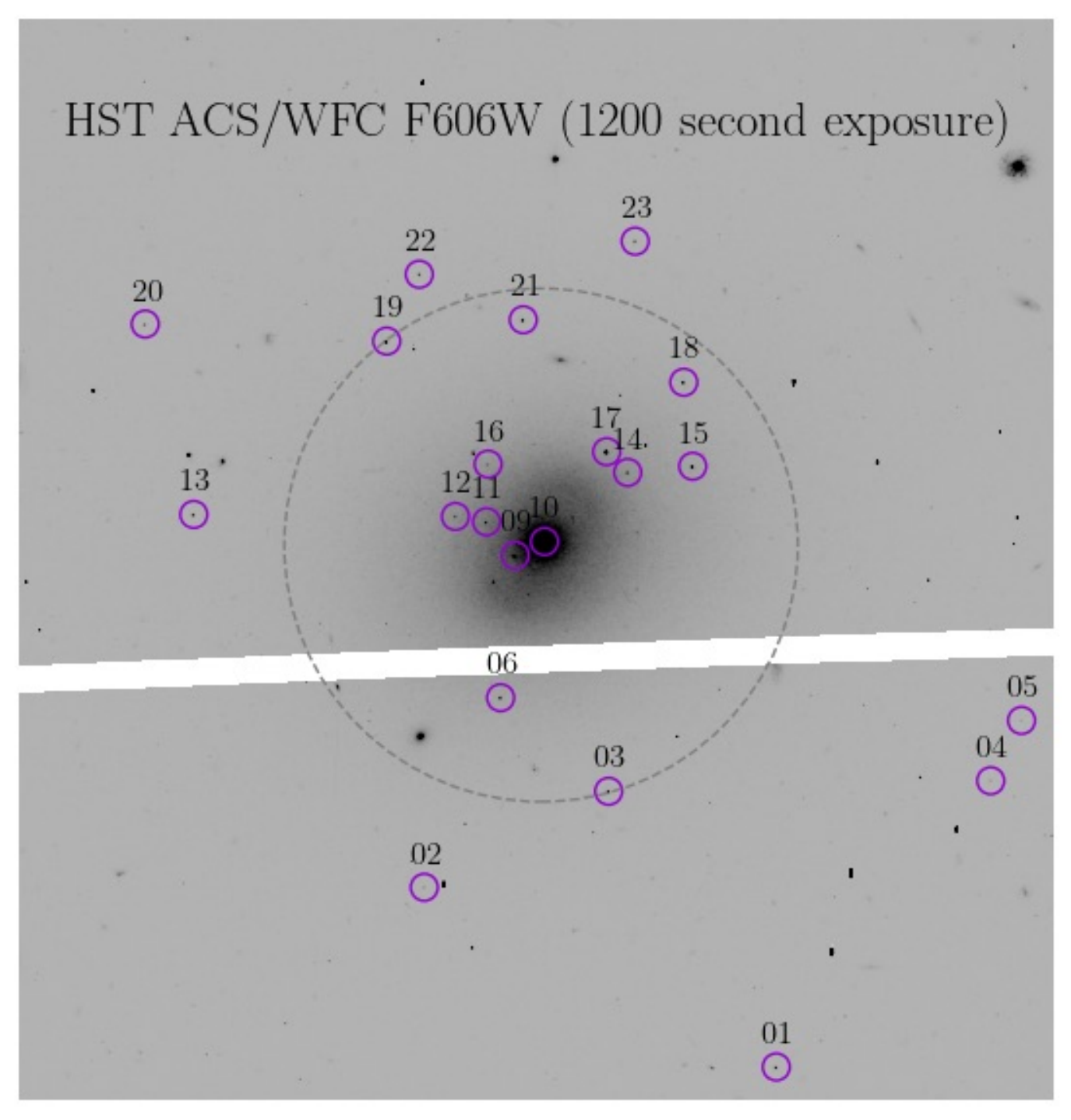}
    \caption{A V-band image of UGC\,7369 taken with the Hubble Space Telescope ACS/WFC (\citealt{Karachantsev_Hubble}). The GCs, selected by \cite{Georgiev_sample1}, are circled in purple and are labeled by their ID as listed in Table \ref{tab:GCs}. The grey dashed circle is the half-light radius of UGC\,7369, $R_\mathrm{e} = 0.9$\,kpc, centered on the galaxy.}
    \label{fig:hubble_cutout}
\end{figure}
Figure \ref{fig:hubble_cutout} shows the HST image of UGC\,7369, with its photometrically-identified GCs circled. After appropriately masking out the clusters, we fit UGC\,7369's surface brightness profile using pymfit \citep{imfit}, and find that it is well-described by a single S\'ersic component with a half-light radius of $R_\mathrm{e} = 15.98" = 0.90$\,kpc at the measured distance of 11.6 Mpc \citep{Georgiev_sample1}, and a S\'ersic index of $n = 1.18$.

In contrast to NGC\,5846-UDG1, which was studied with similar methodology and goals in \citet{UDG1_DF}, UGC\,7369 is a more ``typical'' dwarf galaxy: its GC to stellar mass ratio falls within the intrinsic scatter in the correlation (see e.g. Figure 4 in \citealt{UDG1_GCs}), and it falls nearly exactly on the mass-size relation as well \citep{ElvesI}. As a field galaxy, it also presents a more pristine environment to study the effects of dynamical friction. Additionally, notably, the GC population of UGC\,7369 includes an NSC with a mass of $4\times 10^6\Msun$ and $V-I$ color consistent with the remainder of the GC population, as expected for merger-dominated rather than in-situ NSC growth. As shown in Figure \ref{fig:GCMF}, its mass function also differs somewhat from the universal distribution found from fitting the entire sample: the high mass tail is depleted. Performing a one-sample KS test comparing the non-NSC clusters in UGC\,7369 to the log-normal distribution, we find this difference is statistically significant ($p = 0.048$). As we will discuss, this depletion is an important probe of the system's properties. A summary of relevant parameters for each GC associated with UGC\,7369 used in our analysis is shown in Table \ref{tab:GCs}.

\begin{table}
    \centering
    \begin{tabular}{c|c|c|c} \hline
        ID & $M$ $(10^5\Msun)$ & $V-I$ (mag) & $r_\perp/R_\mathrm{e}$ \\ \hline
        U7369-01 & $0.5\pm0.2$ & $0.9\pm0.10$ & 2.23 \\
        U7369-02 & $0.15\pm0.11$ & $0.8\pm0.16$ & 1.43 \\
        U7369-03 & $1.1\pm0.5$ & $1.0\pm0.10$ & 1.00 \\
        U7369-04 & $0.2\pm0.17$ & $1.0\pm0.15$ & 1.97 \\
        U7369-05 & $0.5\pm0.4$ & $1.3\pm0.16$ & 1.98 \\
        U7369-06 & $0.6\pm0.3$ & $1.0\pm0.10$ & 0.63 \\
        U7369-09 & $3\pm1.3$ & $1.3\pm0.10$ & 0.13 \\
        U7369-10 & $40\pm16$ & $0.82\pm0.08$ & 0.01 \\
        U7369-11 & $0.3\pm0.16$ & $0.9\pm0.13$ & 0.24 \\
        U7369-12 & $0.7\pm0.4$ & $1.15\pm0.13$ & 0.36 \\
        U7369-13 & $0.5\pm0.2$ & $0.9\pm0.10$ & 1.37 \\
        U7369-14 & $0.8\pm0.4$ & $1.1\pm0.11$ & 0.42 \\
        U7369-15 & $1.0\pm0.4$ & $0.93\pm0.09$ & 0.64 \\
        U7369-16 & $0.7\pm0.5$ & $1.2\pm0.15$ & 0.37 \\
        U7369-17 & $1.7\pm0.6$ & $0.90\pm0.08$ & 0.42 \\
        U7369-18 & $1.4\pm0.6$ & $1.1\pm0.10$ & 0.82 \\
        U7369-19 & $1.18\pm0.5$ & $0.92\pm0.09$ & 1.00 \\
        U7369-20 & $0.5\pm0.3$ & $1.0\pm0.12$ & 1.77 \\
        U7369-21 & $0.7\pm0.3$ & $0.9\pm0.10$ & 0.87 \\
        U7369-22 & $0.4\pm0.2$ & $0.94\pm0.12$ & 1.15 \\
        U7369-23 & $0.3\pm0.17$ & $0.85\pm0.10$ & 1.22 \\ \hline
    \end{tabular}
    \caption{Masses, colors, and normalized projected separations from the galaxy's center for the GCs in UGC\,7369, using the values shown in Table 3 of \citet{Georgiev_sample1}. No errors are quoted for the normalized projected separations, so we take them to be $\pm0.01$.}
    \label{tab:GCs}
\end{table}

\begin{figure}
    \centering
    \includegraphics[width=0.47\textwidth]{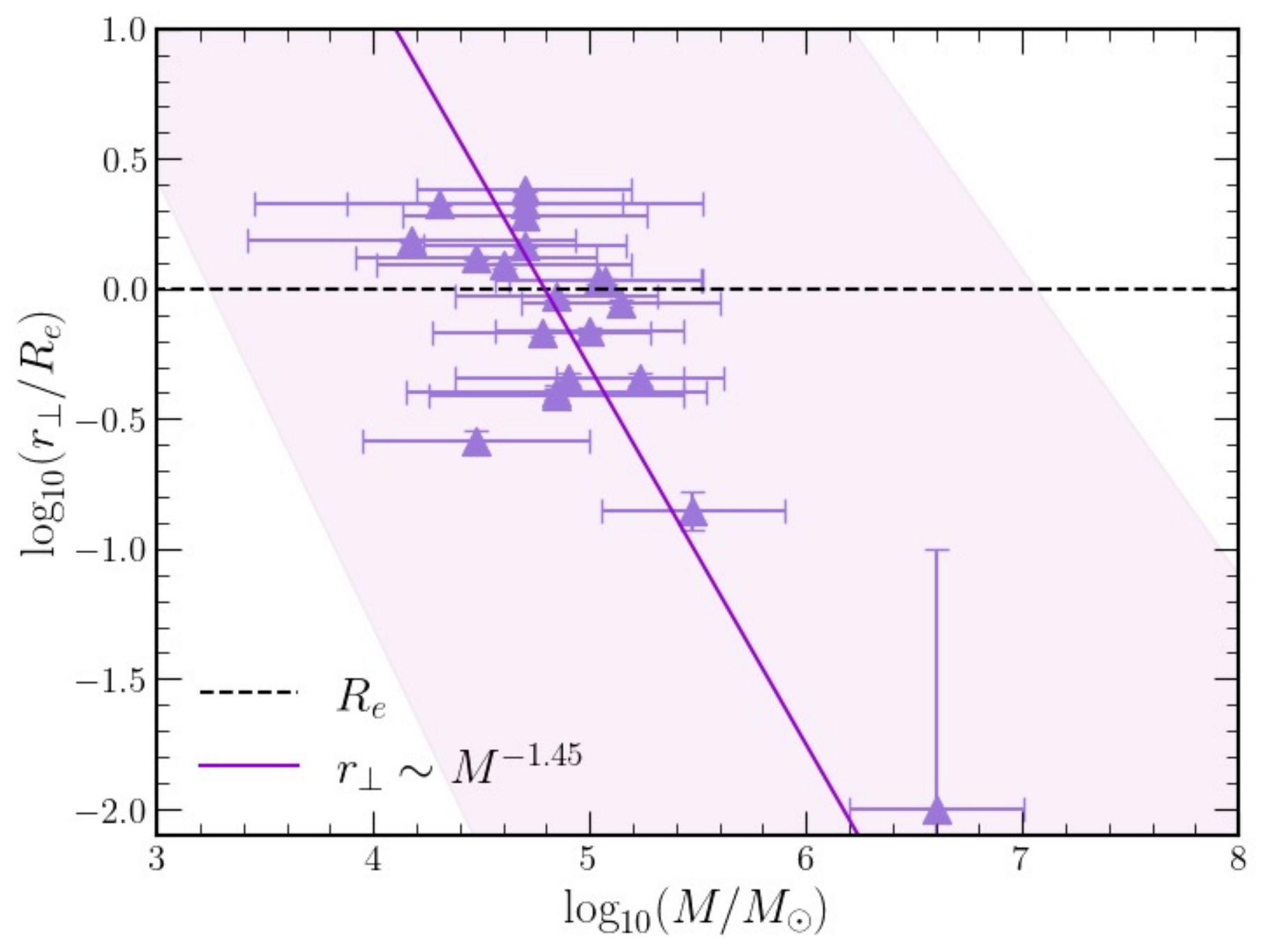}
    \caption{The GC population of UGC\,7369 plotted in a cluster mass-projected radius plane. The dashed black line indicates the measured stellar S\'ersic effective radius, and the solid purple line is a power-law least-squares regression to the data plotted to guide the eye, with the shaded region showing $\pm 1\sigma$ error bars in the fit parameters. The distribution deviates significantly from uniform scatter about $r_\perp = R_\mathrm{e}$ independent of mass, demonstrating mass segregation within this galaxy.}
    \label{fig:observed_segregation}
\end{figure}

Figure \ref{fig:observed_segregation} demonstrates the observed mass segregation among these clusters: in general, clusters with higher masses are found at smaller projected radii. The data are well-described by a power law, here with a slope of $r_\perp \sim M^{-1.45}$, and previous works (e.g. \citealt{UDG1_DF} and references therein) have found that dynamical friction is able to reproduce these trends, as more massive clusters experience stronger friction and thus fall toward the center on shorter timescales. In UGC\,7369, the NSC is clearly visible as the right-most point in the figure, with a mass an order of magnitude larger than any other cluster, and no other clusters are within the inner $\sim 0.1\,\mathrm{kpc}$ of the galaxy; these are the chief features we aim to reproduce in our modeling.

\section{Simulation Design}
\label{sec:sim_design}

To study the effectiveness of dynamical friction in inducing the observed mass segregation effect and contributing to the formation of the NSC, we build a code to evolve GCs forward in time in a dark matter halo potential. Although treating the GCs as extended objects comprised of systems of stars (N-body and ``live,'' as in \citealt{Chowdhury_GC_Nbody}) would be the most physically correct, such a calculation is computationally expensive, and so we instead model the GCs as single particles with semi-analytic treatments of the physics associated with their interactions and internal dynamics. Because we implement the halo as a smooth, static potential, dynamical friction from the halo's constituent particles is not intrinsically captured. The dynamical friction force is therefore implemented as a semi-analytic prescription depending on the mass, position, velocity, and local halo properties of each GC.

The simulation is implemented in the REBOUND N-body code \citep{rebound}, and uses the IAS15 17th order variable symplectic integrator \citep{reboundias15}. The REBOUNDx \citep{reboundx} package is also used in order to implement additional features, such as mass loss. We integrate GC orbits for 10\,Gyr, to study their evolution from the formation of the galaxy to the present day. We use units of kpc-Gyr-$10^5\Msun$, and the timestep is allowed to vary between $10^{-5}$\,Gyr and $10^{-3}$\,Gyr, to balance accurately modeling encounters and mergers with high resolution while remaining computationally efficient. In the absence of dissipative effects like mergers and dynamical friction, we find that energy is conserved to within $\sim 0.1\%$ after 10\,Gyr.

Below, we describe our implementation of the GC particle models (\S\ref{sec:particles}), halo models (\S\ref{sec:halos}), dynamical friction (\S\ref{sec:friction}), and the initialization procedure (\S\ref{sec:ICs}). Finally, \S\ref{sec:simparams} summarizes the input parameters required to generate a model, and \S\ref{sec:modelingUGC} details how we generate models of UGC\,7369 based on the data presented in \S\ref{sec:data}.

\subsection{GC Particle Models}
\label{sec:particles}

In modeling the GCs as single particles, the key effects missing in their dynamics are in their mass and internal energy evolution over time. The dominant contributions to their evolution is mass loss due to stellar evaporation, stellar evolution, and shocks, as well as mergers and interactions between the GCs. Therefore, we prescribe continuous mass loss rates for each particle and a merger criterion based on separation and binding energy in each encounter.

\citet{Shao_massloss} found that mass loss occurs in a very rapid phase during the first $0.5 - 1$\,Gyr, and then more gradually over the next $10$\,Gyr, so assuming that the GCs have a relaxation timescale of at least $0.5$\,Gyr, as found in \citet{UDG1_DF}, we may neglect this initial loss. We use an exponential fit to the post-initial-loss rates from \citet{Shao_massloss},
\begin{equation}
    M_i(t) = M_i^{(0)}\exp(-t/(23\textrm{\,Gyr}))
\end{equation}
and use the time dependent mass of each GC to source a Plummer-softened potential 
\begin{equation}
    \phi_i(\textbf r, t) = \frac{GM_i(t)}{\sqrt{(\textbf{r} - \textbf{r}_i(t))^2 + \epsilon^2}}
\end{equation}
where the index $i = 1\textrm{, }\dots\textrm{, }N_{GC}$ labels the GCs, $\textbf{r}_i(t)$ is each GC's trajectory, and the softening is taken to be $\epsilon = 10$ pc. For the purposes of tracking collisions, each GC is also assigned a particle size $R_i$ given by the size-mass relation in \citet{Brown_Gnedin_Mass_Size}: 
\begin{equation}
    R_i = 2.548\textrm{ pc}\left(\frac{M_i^{(0)}}{10^4\Msun}\right)^{0.242}
\end{equation}

We define a merger between GCs $i$ and $j$ as occurring when the particles pass within $2(R_i + R_j)$ of each other and the relative energy
\begin{equation}
    E_{ij}(\Delta\textbf{r}, \Delta\textbf{v}, t) = \frac{1}{2}\mu(t)(\Delta\textbf{v})^2 + U(\Delta\textbf{r}, t)
\end{equation}
is negative, where $\mu(t) = M_i(t)M_j(t)/(M_i(t)+M_j(t))$ is the reduced mass, $\Delta\textbf{r} = \textbf{r}_i - \textbf{r}_j$ and $\Delta\textbf{v} = \textbf{v}_i - \textbf{v}_j$. Here, following \citet{UDG1_DF}, the relative potential energy $U(\Delta\textbf{r}, t) = \frac{1}{2}\int \textrm{d}^3\textbf{r}(\rho_i\phi_j+\rho_j\phi_i)$ between two Plummer spheres is approximated by
\begin{equation}
    U(\Delta\textbf{r}, t) \approx \frac{-GM_i(t)M_j(t)}{\left(|\Delta\textbf r|^{2.11} + (1.7\epsilon)^{2.11}\right)^{1/2.11}}
\end{equation}
In the event of a merger, we assign the resulting GC a new mass given by the sum of the constituent GC masses, a new size given by applying the size-mass relation to the new mass, a new velocity corresponding to the conservation of momentum, and a new position at the center of mass of the GCs. The total kinetic and orbital potential energy of the GCs is not conserved in this merger process: physically, energy is transferred to the internal energy of the GC, but because we treat GCs as individual particles, this effect is not incorporated in our simulations. While this prescription is certainly an approximation, \citet{Chowdhury_GC_Nbody} demonstrate that modeling mergers in this way reasonably mimics the full N-body internal dynamics of typical merging GCs (e.g. their Figures 7 and 8).  Additionally, because most mergers in our simulations occur in the innermost regions of the galaxy after the GCs have lost energy to the friction, the relative velocities are fairly low, so we expect slightly different merger prescriptions would not substantially change our results.

We find that reasonable variations (of at least up to $\sim 50\%$) in the softening length, particle size, and mass loss timescales do not significantly alter any results.

\subsection{Halo Models}
\label{sec:halos}

To model the galaxy's dark matter halo, we consider both cored models following \citet{Burkert}, with
\begin{equation}
    \label{eq:rho_B}
    \rho_B(r) = \rho_0\left(1 + \frac{r}{r_0}\right)^{-1}\left(1 + \left(\frac{r}{r_0}\right)^2\right)^{-1}
\end{equation}
as well as cuspy NFW models, given by \citet{NFW}:
\begin{equation}
    \label{eq:rho_N}
    \rho_N(r) = \rho_0\left(\frac{r}{r_0}\right)^{-1}\left(1 + \frac{r}{r_0}\right)^{-2}
\end{equation}
where $\rho_0$ and $r_0$ are the scale densities and radii parametrizing the halo. We approximate the properties of the halo as time-independent so that the halos simply contribute to the GC dynamics via a potential term
\begin{equation}
    \phi_h(r) = -G\int_r^\infty \mathrm{d}\tilde{r}\frac{M_h(\tilde{r})}{\tilde{r}^2}
\end{equation}
where $M_h(r) = \int_0^r 4\pi \tilde{r}^2\mathrm{d}\tilde{r}\rho_h(\tilde{r})$ is the enclosed mass of the halo within a radius $r$.

\subsection{Semi-Analytic Dynamical Friction}
\label{sec:friction}

We implement dynamical friction as a velocity-dependent drag force in the dynamics, with each GC experiencing an additional acceleration given by
\begin{equation}
    \mathbf{a_{\mathrm{DF}}}(\mathbf{r}, \mathbf{v})  = -\frac{\mathbf v}{\taudf(M, \mathbf{r}, \mathbf{v})}
\end{equation}
The characteristic friction timescale $\tau_{DF}$ is dependent on the GC's mass, position, speed, and local properties of the halo. For computational simplicity, we approximate the halo distribution function as locally Maxwellian, so that from Chandrasekhar's formula \citep{Chandrasekhar_DF}, we have
\begin{equation}
    \label{eq:taudf}
    \taudf(M, r, v) = \frac{v^3}{2\pi G^2 \log(1 + \Lambda(M, r, v)^2) M C_h(r, v)}
\end{equation}
where 
\begin{equation}
    C_h(r, v) = \rho_h(r)\left(\mathrm{erf}(X) - \frac{2X}{\sqrt{\pi}}e^{-X^2}\right)
\end{equation}
with $X \equiv v/\sqrt{2\sigma_h(r)^2}$ captures the dependence of the friction force on the halo's radial density and velocity dispersion profiles, and the Coulomb logarithm argument $\Lambda = b_{\mathrm{max}}/b_{\mathrm{min}}$ encodes the smallest and largest impact parameters of the problem. Following the parameterizations given in \citet{Petts2015_DF} and \citet{Petts2016_DF}, we use
\begin{equation}
    b_\mathrm{max} = \min\left\{\frac{\rho_h(r)}{\dv{\rho_h}{r}(r)}, r\right\} \textrm{, } b_\mathrm{min} = \frac{GM}{v^2}
\end{equation}
which naturally incorporates the result that friction no longer occurs when the GC's mass becomes comparable to the enclosed mass of the halo.

Furthermore, to model core stalling effects (as explored in e.g. \citealt{Read_core_stalling}, \citealt{Gualandris_stalling}, \citealt{KS_GC_DF}), we cut off dynamical friction when the GC's tidal radius \citep{BT}
\begin{equation}
    \label{eq:r_tidal}
    r_t(M, r) = \left(\frac{GM}{\frac{GM_h(r)}{r^3} - \dv{^2\phi_h}{r^2}(r)}\right)^{1/3}
\end{equation}
exceeds its distance from the center of the halo.
\begin{figure}
    \centering
    \includegraphics[width=0.48\textwidth]{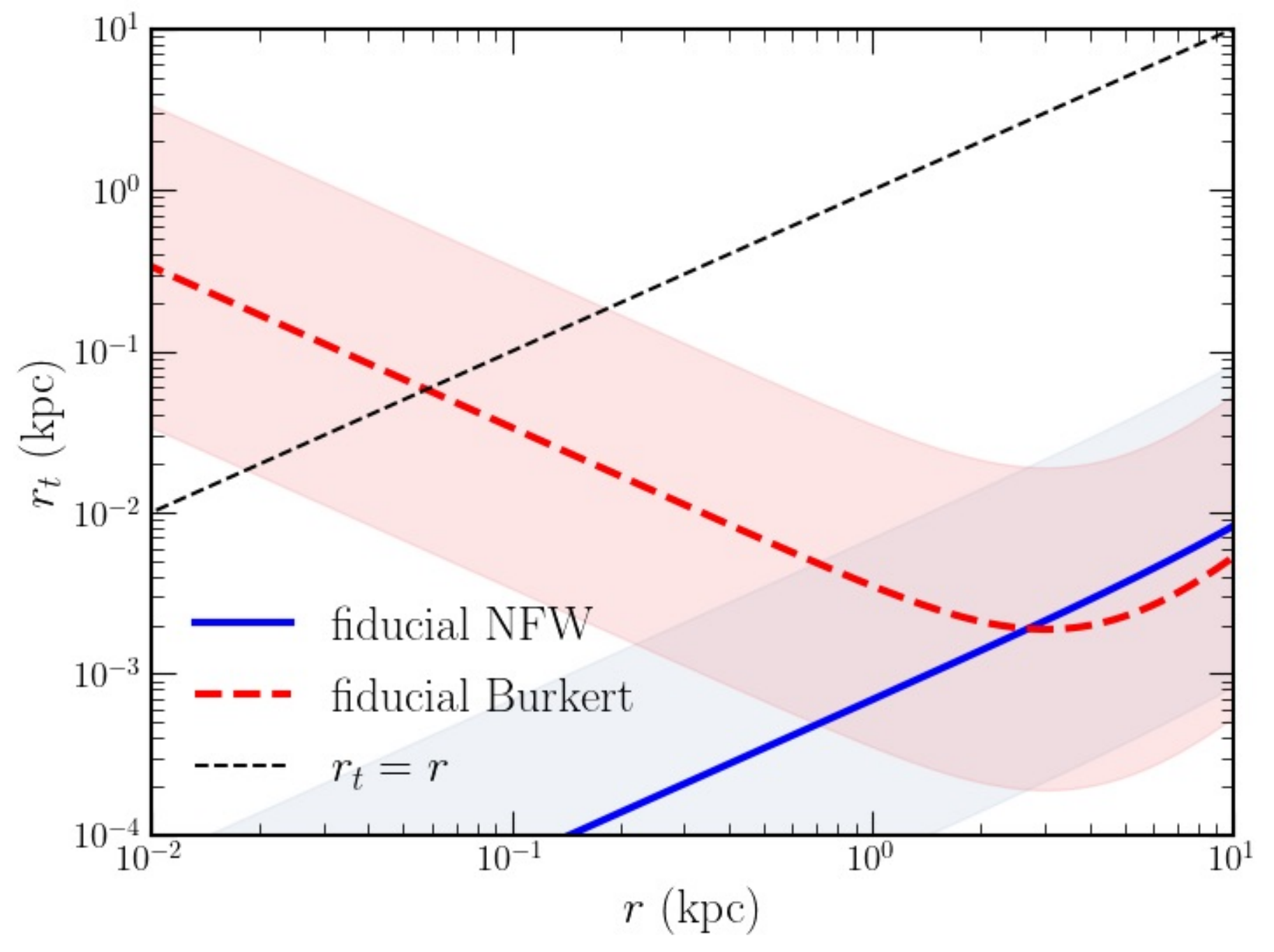}
    \caption{The tidal radius (equation \ref{eq:r_tidal}) as a function of orbital radius for a GC with $M = 10^5\Msun$ in our fiducial NFW and Burkert halos (solid blue and dashed red lines respectively) --- see Table \ref{tab:halos} for details. The shaded regions represent the range of tidal radii for masses between $10^4\Msun$ and $10^6\Msun$. Dynamical friction is cut off when $r_t = r$ (black dashed line), which affects the cored profile, but not the cusp.}
    \label{fig:tidal_radius_plot}
\end{figure}
Figure \ref{fig:tidal_radius_plot} illustrates the effects of this cutoff procedure. At small distances from the halo center, the tidal radius of a GC in a cored density profile increases more rapidly than that of one in a cuspy profile, because in a core, the enclosed mass and derivatives of the potential both tend to zero, whereas in a cusp, the enclosed mass contribution is greater than zero. Thus, in practice, this cutoff has no effect on the cuspy models, while mimicking core stalling in cored halos. So, although both types of halos are treated identically in our simulations, this framework allows for significant differences in the strength and relevance of dynamical friction depending on the presence of a core. This prescription is certainly incomplete, as it is still reliant on the homogeneous Maxwellian approximation for the halo distribution function and lacks effects such as dynamical buoyancy \citep{Banik_vdB_friction} in the core, but has been demonstrated to reasonably match N-body results \citep{Petts2016_DF}.

\subsection{Initial Conditions}
\label{sec:ICs}
Each GC is initialized with a mass $M_i^{(0)} = M_i^*\exp(10/23)$, where $M_i^*$ is drawn from the log-normal fit to the present-day observed mass function displayed in Figure \ref{fig:GCMF}, and a correction is applied such that the final total GC mass $M_{GC}$ matches observations after 10\,Gyr of mass loss. We draw GC masses until the difference between the total drawn mass and the target $M_{GC}$ is on average within 1 draw of the total specified GC mass, i.e. until $M_{GC} - \sum_i M_i^* < 10^{5}\Msun$, and then draw one additional GC mass. In this way, each GC mass is drawn from the same universal distribution, and the total mass in GCs on average matches the observed present-day total, though the initial number of GCs and total mass in GCs in each realization may differ.

Initial orbital radii are drawn from the distribution $P(r) = 4\pi r^2\nu(r)$ where the radial density profile $\nu(r)$ is given by \citet{PrugnielSimien},
\begin{equation}
    \nu(r) \sim \left(\frac{r}{R_\mathrm{e}}\right)^{p(n)}\exp\left(-b(n)\left(\frac{r}{R_\mathrm{e}}\right)^{1/n}\right)
\end{equation}
This functional form is intended to provide a simple analytic approximation to a deprojected S\'ersic profile of effective radius $R_\mathrm{e}$ and S\'ersic index $n$, where $p(n)$ and $b(n)$ are numerical coefficients depending on the S\'ersic index (equations A3a and B5 of \cite{PrugnielSimien}), and the distribution is normalized so that $\int P(r) dr = 1$. For reference, using the fitted $n = 1.18$ for UGC\,7369, we find $p(1.18) \approx 0.52$ and $b(1.18) \approx 2.04$. For the range of $R_\mathrm{e}$ and $n$ values considered in our UGC\,7369 models, the projected density differs from a true S\'ersic by a typical value of $\lesssim 1\%$, with the inner regions of interest matching to $\lesssim 0.1\%$. A 3D position is then assigned to each GC by drawing spherical angle coordinates $\theta$ and $\varphi$ from uniform distributions over $[0, \pi]$ and $[0, 2\pi)$ respectively.

We choose initial velocities in such a way that in the absence of friction and mergers, the radial distribution remains in a (statistical) steady state, which entails using an ergodic distribution function $f(\textbf{r}, \textbf{v}) = f(\Psi(\textbf{r})-\frac{1}{2}\textbf{v}^2)$, where by convention $\Psi(\textbf{r}) = -\phi_h(r)$, and we normalize so that $\int \textrm{d}^3\textbf{r}\textrm{d}^3\textbf{v} f = 1$. Such a distribution function may be calculated numerically using the Eddington formula and integrating by parts \citep{BT}:
\begin{align}
    \label{eq:Eddington}
    f(\mathcal{E}) = \frac{1}{\sqrt{8}\pi^2}\bigg(
    \frac{1}{\sqrt{\mathcal{E}}}\dv{\nu}{\Psi}\bigg|_0 + & 2\sqrt{\mathcal{E}}\dv{^2\nu}{\Psi^2}\bigg|_0 \\ \nonumber & + 2\int_0^\mathcal{E}\textrm{d}\Psi \dv{^3\nu}{\Psi^3}\sqrt{\mathcal{E}-\Psi}\bigg)
\end{align}
where we invert the potential to write $\nu = \nu(r(\Psi))$. For the Burkert and NFW halo potentials we consider, $\Psi\to 0$ only as $r\to\infty$, so the first two boundary terms are zero and we are just left with the final integral to calculate numerically. Equation \ref{eq:Eddington} guarantees that the distribution function is positive, and we have checked that the resulting $f(\mathcal{E})$ values are numerically stable. We then calculate the conditional distribution of velocities at each position: for a GC initialized at $r$, the initial speed is drawn from
\begin{equation}
    P(v|r) = 4\pi v^2 \frac{f\left(\Psi(r) - \frac{1}{2}v^2\right)}{\nu(r)}
\end{equation}
and we again draw spherical angle coordinates $\theta_v$ and $\varphi_v$ from uniform distributions over $[0, \pi]$ and $[0, 2\pi)$, respectively, to determine the direction of the initial velocity.

\subsection{Summary of Parameters and Expectations}
\label{sec:simparams}

One dynamical model can be fully specified by the following 6 parameters:
\begin{itemize}
    \item the type of the dark matter halo (cusp or core), which influences the qualitative properties of the dynamical friction the GCs experience
    \item halo density profile parameters $(\rho_0, r_0)$, which govern the details of the potential in which the GCs move
    \item S\'ersic profile parameters $(R_\mathrm{e}, n)$, which determine the initial radial distribution of the GCs
    \item the total mass in GCs $M_{GC}$, which implicitly sets the typical number of GCs through random draws from the universal mass function
\end{itemize}

In general, we anticipate the GC population in both types of halos to undergo mass segregation, because according to equation \ref{eq:taudf}, the friction timescale approximately scales as $\tau \sim 1/M$, so more massive clusters will fall to the center of the halo more quickly. However, because of core stalling, dynamical friction shuts off near the centers of cored halos, so the GCs do not sink all the way through the core. In contrast, in cuspy models, the most massive GCs are free to sink much deeper into the potential well of the halo, and are more likely to merge with other GCs to form an NSC there. We expect mergers to mainly take place in the center of the halo, where the GC sizes become larger relative to their orbital radii, and so the extent of the inspiral allowed by the halo can significantly alter the final GC mass function.

In comparison to the similar analysis in \citet{UDG1_DF}, our simulation code differs in two important ways. First, our prescription of dynamical friction is updated: we estimate the minimum and maximum impact parameters differently, and also model how dynamical heating and core stalling shut off friction in a way that depends on the mass of each individual GC and local characteristics of the halo, instead of a single fixed cutoff radius. Second, and most importantly, instead of initializing the GC populations with masses exactly matching the observed present-day distribution corrected for mass loss effects, we allow for greater flexibility in the initial number of GCs and their masses by using random draws from the universal mass function described in \S\ref{sec:data}.

In this way, the final distribution of GC masses becomes another observable that we use to constrain properties of the halo, since our initial distribution is not fixed to observations. As a simple validation test for our simulation code and new friction prescription, we confirm that for parameter choices matching UDG1, we recover similar friction timescales (presented in Figure 4 of \citealt{UDG1_DF}) and segregation slopes (presented in Figure 5 of \citealt{UDG1_DF}) if we use the same ``exact'' initial mass distribution matched to present observations.

\subsection{Modeling UGC\,7369}
\label{sec:modelingUGC}

\begin{figure*}
    \centering
    \includegraphics[width=0.98\textwidth]{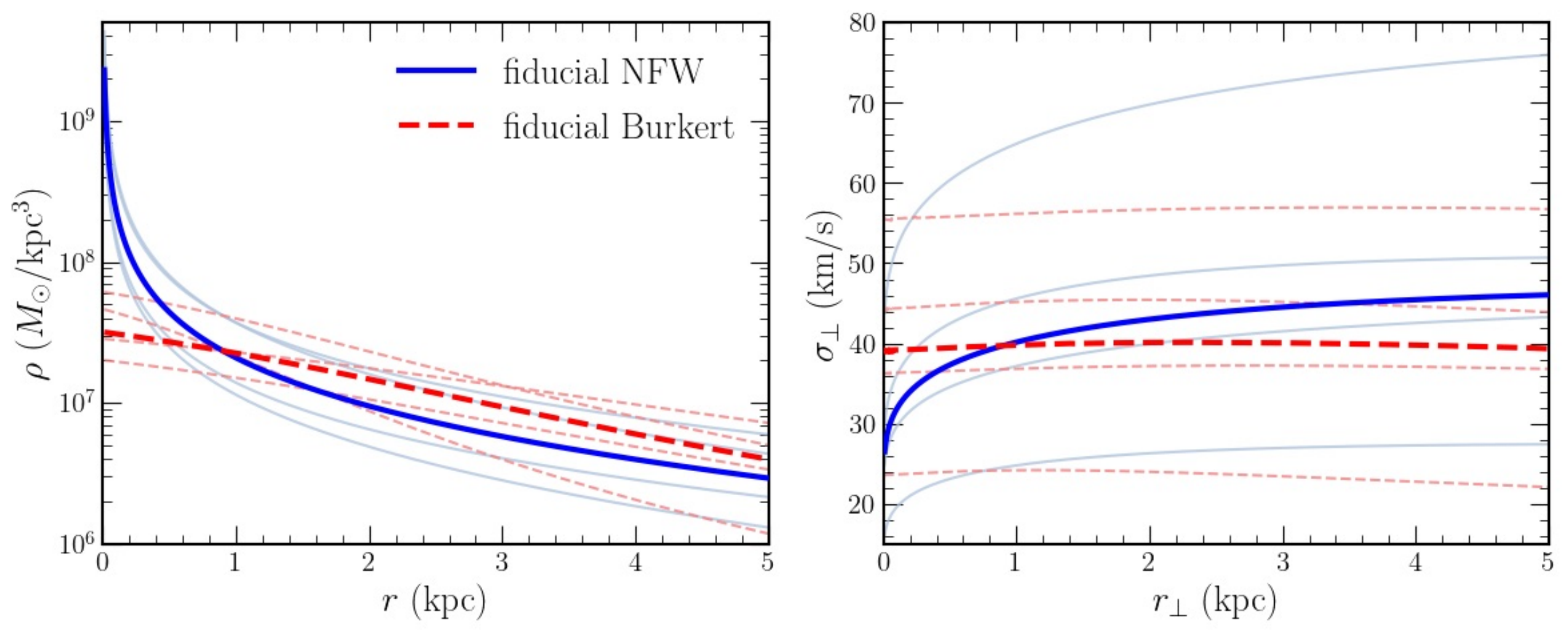}
    \caption{Radial density and projected velocity dispersion profiles for the fiducial NFW and Burkert halos (solid blue and dashed red lines respectively), as well as for NFW and Burkert halos with other varied parameter choices (solid light blue and dashed light red lines respectively); see \S\ref{sec:grid} for details.}
    \label{fig:halo_models}
\end{figure*}

\begin{figure*}
    \centering
    \includegraphics[width=0.98\textwidth]{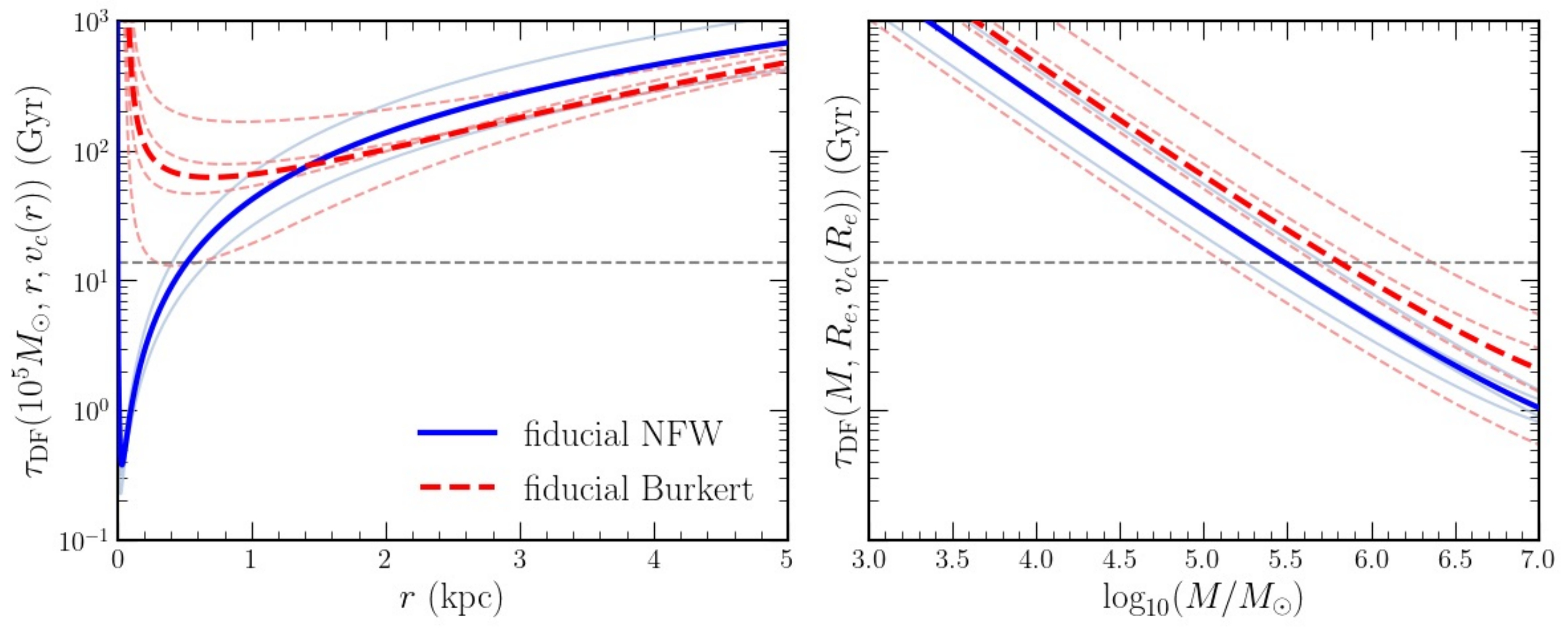}
    \caption{The dynamical friction timescales for the fiducial NFW and Burkert halos (solid blue and dashed red lines respectively), as well as for NFW and Burkert halos with the same varied parameter choices as in Figure \ref{fig:halo_models} (solid light blue and dashed light red lines respectively). The left panel shows the timescale at a fixed typical GC mass of $10^5\Msun$ at varied radius, and the right panel shows the timescale at fixed $r = R_\mathrm{e}$ but with varied GC mass. In both panels, it is assumed that the GC is traveling at the circular velocity at its current radius. For reference, the grey dashed line indicates the Hubble time.}
    \label{fig:taudf}
\end{figure*}

Having validated the simulation code, in order to test the effectiveness of cored and cuspy halos in reproducing observed features of the GC population of UGC\,7369, we choose a fiducial set of simulation parameters as follows. We take the S\'ersic parameters for GC initialization to match the fit to the photometry presented in \S\ref{sec:data}, $R_\mathrm{e} = 0.9$\,kpc and $n = 1.18$. We also match the total GC mass to the observed total $M_{GC} = 6\times 10^6\Msun$, which amounts to the assumption that all the observed GCs are at least 10\,Gyr old. Because the most massive observed GCs have colors reasonably consistent with such an age, and they dominate the total $M_{GC}$, this is a sensible approximation to the system. The fiducial NFW halo mass is determined by using the stellar-halo mass relation
\begin{equation}
    \label{eq:SHMR}
    \log_{10}\left(M_{200}/M_1\right) = \frac{1}{\alpha}(\log_{10}\left(M_\star/M_1\right) - \log_{10}(\epsilon))
\end{equation}
where $M_1 \equiv 10^{12.5}\Msun$, $\alpha \equiv 1.74$, and $\log_{10}(\epsilon) \equiv -1.70$ (Danieli et al., in prep). Taking $M_\star = 1.72\times 10^8\Msun$, this yields a fiducial $M_{200} = 1.06\times 10^{11}\Msun$. Next, we use the $M_{200}$-concentration scaling relations calculated in \citet{Diemer_Joyce} to determine a fiducial concentration of $c = 5$ (corresponding to our fiducial halo mass at approximately $z = 1$) and use these values to determine the scale density and radius. Finally, the fiducial Burkert halo parameters are determined by matching the fiducial NFW halo's density and projected velocity dispersion at $R_\mathrm{e}$. Figure \ref{fig:halo_models} shows the radial density and projected velocity dispersion profiles for the fiducial parameter choices.
The dynamical friction timescales associated with each halo is pictured in Figure \ref{fig:taudf} as a function of both mass and radius, assuming circular orbits for reference.

As demonstrated in Figure \ref{fig:taudf}, regardless of halo type and parameter choice, the friction timescale is only less than a Hubble time (shown by the horizontal dashed grey line) for the innermost radii (left panel) and largest cluster masses (right panel), so in each simulation, we expect results to vary depending on the specific randomly drawn GC masses and initial radii of the largest GCs. In order to ensure that our conclusions are independent of these initial conditions, we carry out 50 simulations in each halo, and study the results after averaging over these realizations. For each simulation realization of initial conditions, in addition to tracking the full 6D phase space trajectory and merger history of each GC in 0.5\,Gyr output timesteps, we model projected quantities by ``re-observing'' the final state of the galaxy: we randomly uniformly draw angles $\theta_0 \in [0, \pi]$ and $\varphi_0 \in [0, 2\pi)$ to define an isotropic distribution of possible lines of sight $\hat{z}_0 = (\sin\theta_0\cos\varphi_0, \sin\theta_0\sin\varphi_0, \cos\theta_0)$, and calculate
\begin{equation}
    r_{\perp, i} = \left\langle\sqrt{r_i^2 - \left(\textbf{r}_i\cdot\hat{z}_0\right)^2}\right\rangle
\end{equation}
where the index $i$ labels a given GC, $\textbf{r}_i$ is its 3D position, and the average is taken over 100 different viewing directions $\{\hat{z}_0\}$. After this post-processing step, our final data for each simulation in each halo include the masses and projected radii of each GC in steps of 0.5\,Gyr, with errors on the projected radii calculated as standard deviations between the 100 projections.

\section{Results}
\label{sec:results}

From these data, we compare the final simulated GC populations to the UGC\,7369 observations (\S\ref{sec:UGC7369}) by calculating the mass segregation slope of the linear fit to the points in $(\log_{10}(M/\Msun), \log_{10}(r_\perp/R_\mathrm{e})$ space similar to \cite{UDG1_DF}, and also comparing features of the GC mass functions.

\subsection{GC Mass Segregation}

\begin{figure*}
    \centering
    \includegraphics[width=0.98\textwidth]{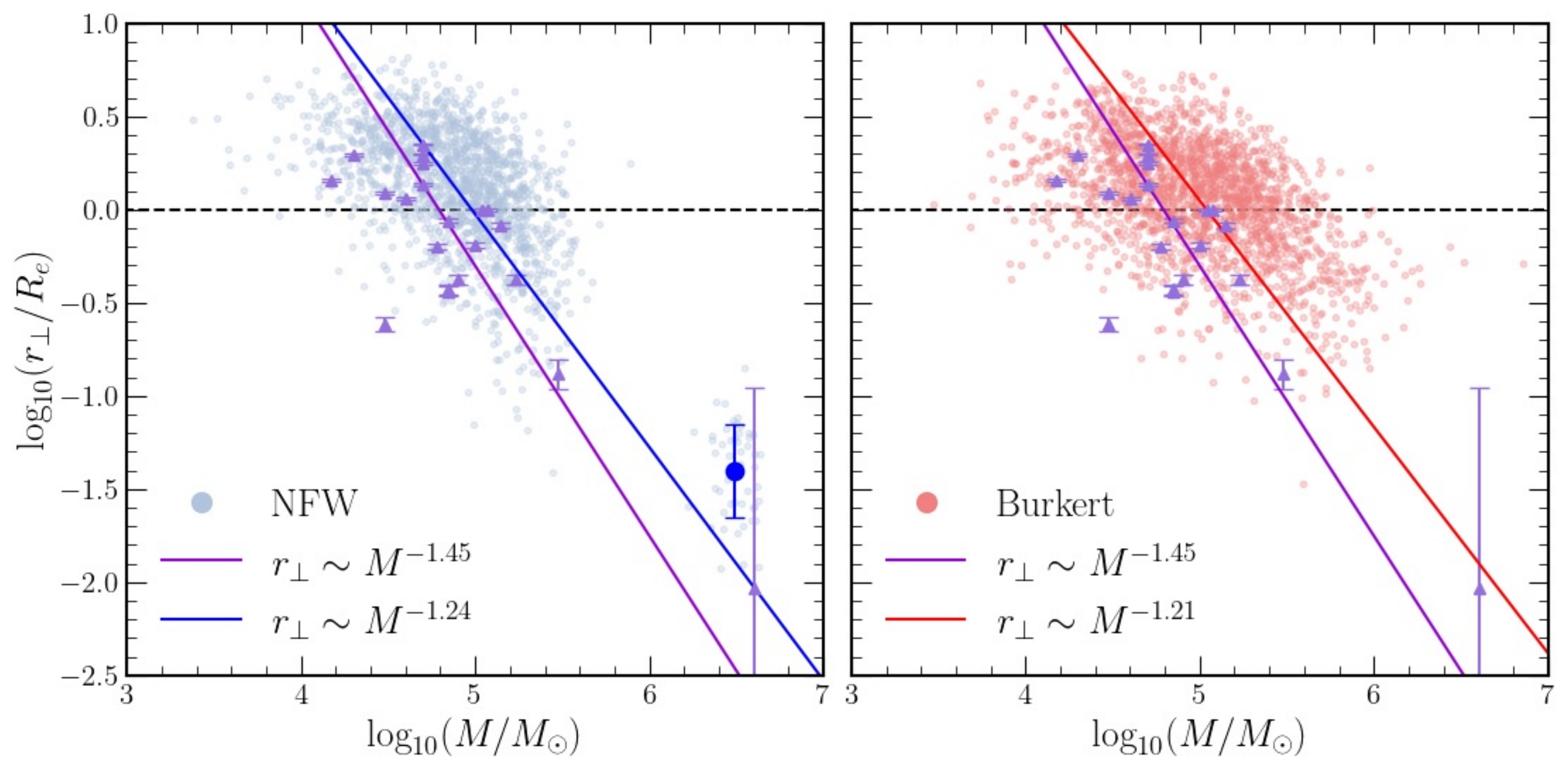}
    \caption{An example of the simulation results for the NFW (left) and Burkert (right) halos. Each of the lighter colored blue or red points respectively is one GC after 10\,Gyr of evolution, and points from all 50 realizations are shown, in addition to the observed GCs in purple. Linear regressions to the points demonstrating the segregation are also included, and the dark blue point on the NFW panel shows the mean properties of the NSC. We note that the linear fit for the NFW model is not significantly changed if the NSC points are excluded. The black dashed line indicates the radius $R_\mathrm{e}$ around which GCs are initialized, independent of mass. Mass errors on the observed points and projection errors on the simulated points are suppressed for clarity.}
    \label{fig:rperp_vs_M}
\end{figure*}

The resulting masses and projected radii of the simulated GC populations after 10\,Gyr of evolution in the fiducial NFW and Burkert halos are displayed in the left and right panels of Figure \ref{fig:rperp_vs_M}, respectively.
We calculate the segregation slope by using least squares regression to fit GCs from all 50 realizations of the simulation at once, and find that both halo models yield similar results, so it is challenging to conclude which halo yields the more realistic fit, although we see that dynamical friction certainly has the capacity to fully account for the observed segregation, as found in \cite{UDG1_DF}.

However, the key striking difference between the halo results is that the NFW halo produces a massive GC near the center via mergers, while the Burkert halo simulations never result in a clearly identifiable NSC (i.e. a substantially more massive cluster also being the closest to the center of the halo). We characterize the properties of the typical NSC formed in the cuspy halo by averaging the masses and projected radii of the GCs with $M > 10^{6.2}\Msun$, resulting in the dark blue point shown in the bottom right of the left panel. We find that the properties of the typical simulated NSC match those of the observed one well within the observational error in both mass and projected separation.

\subsection{The GC Mass Function}

\begin{figure*}
    \centering
    \includegraphics[width=0.98\textwidth]{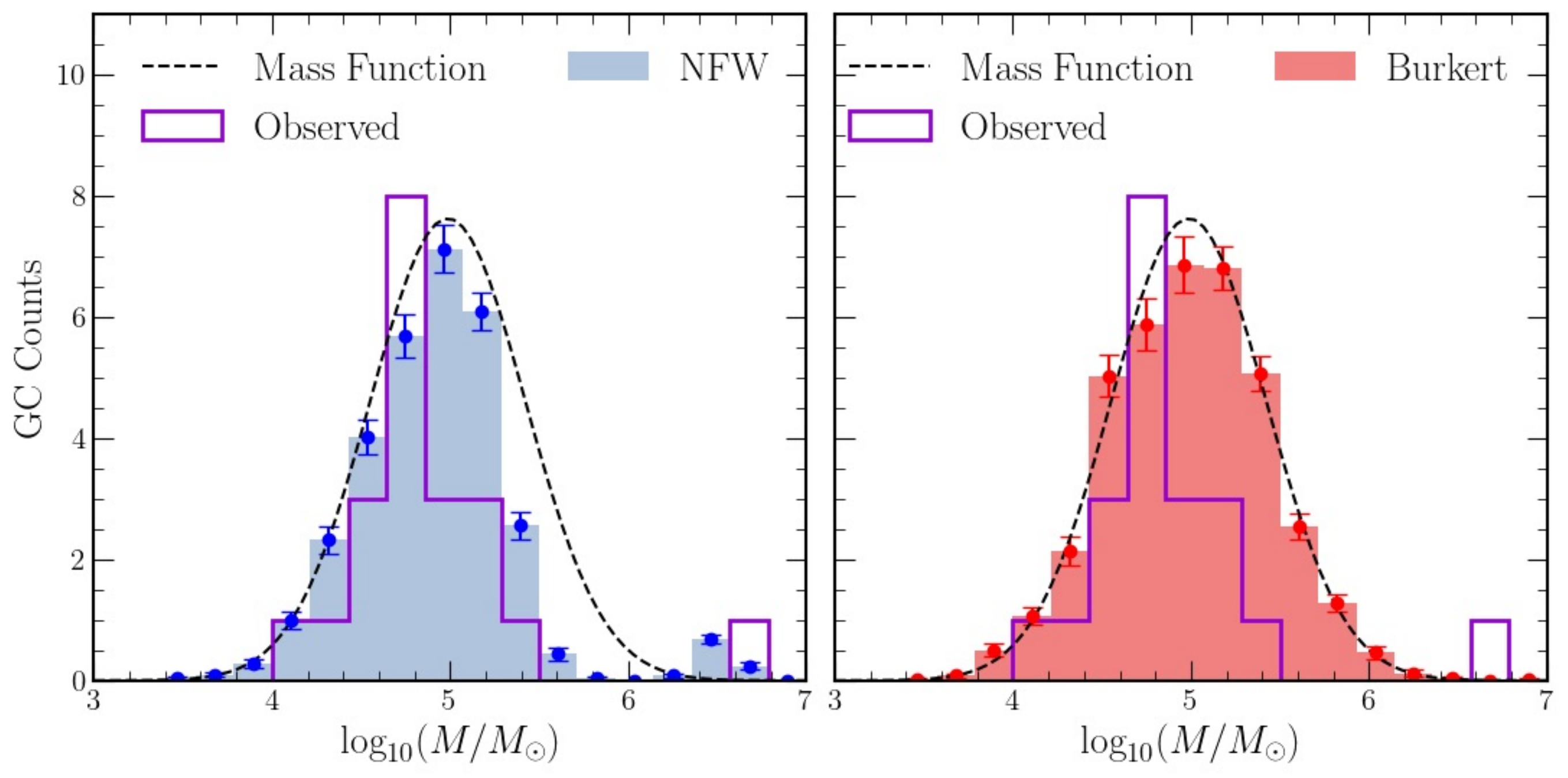}
    \caption{The GC mass function after 10\,Gyr in the fiducial halo models of each type. The counts in each bin have been averaged over the 50 realizations, and the error bars are calculated from the $\pm 1\sigma$ variation in each bin between realizations. The NFW model better matches the observations (shown by the purple histograms), both in the occurrence of an NSC and in the dearth of clusters with masses near $10^6\Msun$. In both cases, $\langle M_{GC}\rangle = (6.1\pm 0.2)\times 10^6\Msun$, but more of the cluster mass is in the NSC mass bins in the NFW halo histogram. The black dashed curve superimposes the GC mass function from Figure \ref{fig:GCMF}; the Burkert distributions are essentially unchanged after 10\,Gyr.}
    \label{fig:simulated_gcmf}
\end{figure*}

The distinction between the ability of each halo to precipitate GC mergers and form an NSC is even more clearly apparent in the GC mass functions resulting from the simulations after 10\,Gyr. Figure \ref{fig:simulated_gcmf} shows the projection of the data from Figure \ref{fig:rperp_vs_M} onto just the mass dimension, with averaging done over the 50 realizations and binning performed using the same bins as the universal mass function from Figure \ref{fig:GCMF}.
While the average total GC mass across the realizations is identical in both halos, matching the observations, due to a higher frequency of mergers between clusters, the NFW halo ended up with seven fewer GCs present on average than the Burkert halo. Comparing with the observations, the cuspy halo results more closely match the depletion of UGC\,7369's high mass clusters relative to the universal mass function, while the cored halo results show the universal mass function draws essentially unaltered. The origin of the depletion is the higher merger rate among those GCs with $M \sim 10^6\Msun$: as shown in Figure \ref{fig:taudf}, the typical friction timescale for these GCs to migrate inward from $R_\mathrm{e}$ is just $2-5$\,Gyr, so in the cuspy halo, they have a much higher tendency to reach the innermost $0.05$\,kpc of the halo, where mergers are significantly more likely to take place due to the finite size of the GCs.

Importantly, the presence of an NSC together with the depletion of the high mass GCs in observations allows for two main conclusions: first, that UGC\,7369's NSC may plausibly have been formed through mergers, and second, that its dark matter halo is cuspy, rather than cored.

\subsection{Dependence on Halo Parameters}
\label{sec:grid}

To test the robustness of the correlation between NSC formation by GC mergers and the presence of a cuspy halo, we carried out simulations for 4 additional NFW and 4 additional Burkert halos, each with parameter variations chosen to represent uncertainties in the fiducial choices and to test the limits of the effects of friction. Each model's scale density $\rho_0$ and radius $r_0$ values are summarized in Table \ref{tab:halos}.

\begin{table*}
    \centering
    \begin{tabular}{c|c|c|c|c|c}
        ID & $\rho_0$ $(10^5\Msun / \textrm{kpc}^3)$ & $r_0$ (kpc) & $M_h(R_\mathrm{e})$ ($10^8\Msun$) & $\sigma_{\perp}(R_\mathrm{e})$ (km/s) & Notes \\
        \hline
        NFW-1 & 11.82 & 19.54 & 1.11 & 39.72 & fiducial stellar mass, $c = 5$ \\
        NFW-2 & 5.35 & 27.91 & 0.73 & 36.76 & $c = 3.5$ \\
        NFW-3 & 35.48 & 12.21 & 2.00 & 45.12 & $c = 8$ \\
        NFW-4 & 11.82 & 33.41 & 1.94 & 64.18 & $5\times$ fiducial stellar mass \\
        NFW-5 & 11.82 & 11.43 & 0.62 & 24.54 & $0.2 \times$ fiducial stellar mass \\
        Burkert-1 & 318.76 & 3.36 & 0.78 & 39.72 & matching NFW-1's $\rho(R_\mathrm{e})$ \& $\sigma_\perp(R_\mathrm{e})$ \\
        Burkert-2 & 201.66 & 3.92 & 0.51 & 36.78 & matching NFW-2's $\rho(R_\mathrm{e})$ \& $\sigma_\perp(R_\mathrm{e})$ \\
        Burkert-3 & 614.45 & 2.74 & 1.42 & 45.10 & matching NFW-3's $\rho(R_\mathrm{e})$ \& $\sigma_\perp(R_\mathrm{e})$ \\
        Burkert-4 & 285.30 & 5.04 & 0.75 & 56.08 & $1.5\times$ Burkert-1's $r_0$ but same $\rho(R_\mathrm{e})$ \\
        Burkert-5 & 463.65 & 1.68 & 0.87 & 24.17 & $0.5\times$ Burkert-1's $r_0$ but same $\rho(R_\mathrm{e})$ \\ \hline
    \end{tabular}
    \caption{A summary of the halo models used in this work. From left to right, the columns list an ID for each halo, scale density and radius parameters, enclosed mass at $R_\mathrm{e}$, projected velocity dispersion at $R_\mathrm{e}$, and a description of the parameter choice.}
    \label{tab:halos}
\end{table*}

NFW-2 and NFW-3 are halos of varied concentration at fixed $M_{200}$, which may be interpreted as a test of the static halo approximation by initializing the halo at different redshifts: a model with $c = 3.5$ corresponds to approximately $z = 2$, and $c = 8$ corresponds to approximately $z = 0.5$. NFW-4 and NFW-5 probe the sensitivity of our conclusions to variations in the total halo mass: they both represent fixed concentrations, but $M_{200}$ is varied by using the stellar-halo mass relation of equation \ref{eq:SHMR} with the input stellar masses artificially varied by factors of $5$ and $1/5$ respectively.

Burkert-2 and Burkert-3 are cored analogues of NFW-2 and NFW-3, with parameters determined by matching the density and projected velocity dispersion in the same way as the fiducial Burkert halo. Furthermore, because the central core is the main feature that reduces the effect of dynamical friction, we consider Burkert profiles that match the same $\rho(R_\mathrm{e})$ as the fiducial case, but include $r_0$ parameters varied by factors of $1.5$ and $0.5$ in Burkert-4 and Burkert-5 respectively. These numbers are chosen to approximately match the projected velocity dispersion at $R_\mathrm{e}$ of NFW-4 and NFW-5, respectively.

Results from all these halos are shown in Table \ref{tab:results_summary}. Although the values of the segregation slope shift slightly, they remain approximately within $\pm 1\sigma$ of each other, and it remains true that NSCs only form in the cuspy halos, where the mass function is always depleted as in the left panel of Figure \ref{fig:simulated_gcmf}. Even in the Burkert-4 halo where the core size is significantly reduced and the typical friction timescale is shorter than any NFW model for all radii $\gtrsim 0.5$\,kpc, because core stalling is still present, there is only $\lesssim$1 merger per 10\,Gyr. In contrast, even in the NFW-4 halo with the highest mass and longest typical friction timescale, at least 3 mergers occur per 10\,Gyr, and an easily identifiable NSC always eventually forms.

\begin{table*}
    \centering
    \begin{tabular}{c|c|c|c|c|c|c}
        ID & Mass Segregation Slope & $\langle N_{GC}\rangle$ & $\langle N_{\textrm{mergers}} \rangle$ & $\langle M_{NSC} \rangle$ $(10^5\Msun)$ & $\langle r_{\perp\textrm{, NSC}}/R_\mathrm{e} \rangle$ & Notes \\
        \hline
        Observations & $-1.4\pm0.3$ & 21 & N/A & $40\pm16$ & $0.010\pm0.010$ & N/A \\
        NFW-1 & $-1.25\pm0.03$ & $31\pm7$ & $7\pm3$ & $32\pm7$ & $0.04\pm0.02$ & fiducial \\
        NFW-2 & $-1.26\pm0.03$ & $33\pm8$ & $7\pm3$ & $32\pm7$ & $0.04\pm0.03$ & lower concentration \\
        NFW-3 & $-1.30\pm0.03$ & $30\pm6$ & $7\pm3$ & $34\pm8$ & $0.03\pm0.017$ & higher concentration \\
        NFW-4 & $-1.24\pm0.03$ & $33\pm6$ & $5\pm3$ & $28\pm8$ & $0.04\pm0.017$ & higher stellar mass \\
        NFW-5 & $-1.18\pm0.03$ & $28\pm6$ & $9\pm3$ & $37\pm6$ & $0.04\pm0.017$ & lower stellar mass \\
        Burkert-1 & $-1.21\pm0.04$ & $38\pm9$ & $0.10\pm0.3$ & N/A & N/A & matching NFW-1 \\
        Burkert-2 & $-1.36\pm0.06$ & $39\pm8$ & $0\pm0.14$ & N/A & N/A & matching NFW-2 \\
        Burkert-3 & $-1.13\pm0.03$ & $37\pm8$ & $0\pm0.14$ & N/A & N/A & matching NFW-3 \\
        Burkert-4 & $-1.36\pm0.06$ & $38\pm8$ & 0 & N/A & N/A & larger core radius \\
        Burkert-5 & $-1.21\pm0.03$ & $38\pm8$ & $0.5\pm0.6$ & N/A & N/A & smaller core radius \\ \hline
    \end{tabular}
    \caption{A summary of simulation results including the slope of a best-fit line to the binned $\log_{10}(r_\perp/R_\mathrm{e})$ vs. $\log_{10}(M/\Msun)$ data, the average final number of GCs, the average number of mergers, the average NSC mass, and the average projected separation of the NSC. Averages are taken over the 50 realizations of each halo model, and the quoted error ranges are $\pm1\sigma$.}
    \label{tab:results_summary}
\end{table*}

Overall, the dynamical friction timescale is not a strong function of the halo concentration. The total mass of the halo matters more, since a higher density at a fixed radius leads to a longer typical friction timescale for a circular orbit due to the $\taudf \sim v^3/\rho$ dependence. As previously described, it is only the high mass tail of the GC population with $M \gtrsim 10^{5.5}\Msun$ that is able to migrate inward significantly, and this migration will be present across a very wide range of halo parameters. However, in a cored halo, core stalling is a strong enough effect to prevent the complete inspiral of even these GCs, as at radii within $\lesssim 0.1$\,kpc, the friction timescale is orders of magnitude larger than the Hubble time. In contrast, for a cuspy halo, the stalling effect is significantly reduced, and when these massive GCs fall in deeper, they are correspondingly more likely to merge and decrease the friction timescale even further, leading to the growth of the NSC.

\subsection{Evolution of the Nuclear Star Cluster}
\label{sec:NSC_evolution}
\begin{figure*}
    \centering
    \includegraphics[width=0.98\textwidth]{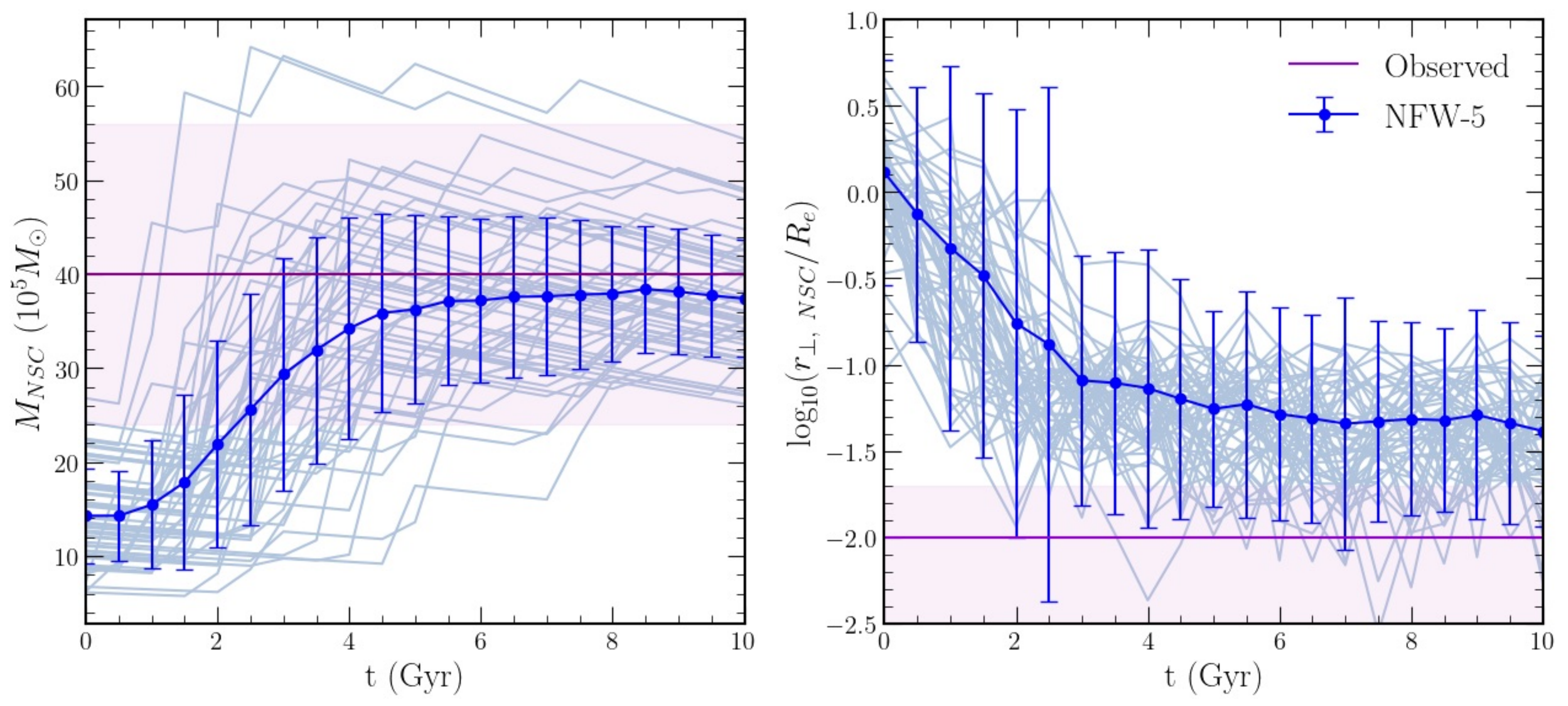}
    \caption{The evolution of the mass and projected radius of the cluster that becomes the NSC over time in the NFW-5 halo model. The light blue lines represent individual realizations, and the darker blue line shows the average value and $\pm1\sigma$ error bars every 0.5\,Gyr. The purple line indicates the observations, with the shaded region around it representing $\pm 1\sigma$ errors.}
    \label{fig:NSC_evolution}
\end{figure*}

Finally, now that we have established the robustness of the connection between the presence of the NSC and the cuspy halo, we may also study properties of the NSC as it evolves and undergoes mergers. Figure \ref{fig:NSC_evolution} shows the mass and projected separation radius of the NSC progenitor in each of the 50 realizations of the lower stellar mass NFW-5 model, which led to the closest match to the observed NSC properties. We identify the progenitor as the highest mass cluster that participates in the first merger near the halo's center, and this is typically the GC with the largest initial mass.
We see that the NSC first begins to form at $\sim 1-2$\,Gyr, consistent with the shortest dynamical friction timescales typically present in the system, and is able to reach its final mass and projected radius by $\sim 4$\,Gyr. Although some smaller mergers continue to occur after this time, they are mostly balanced out by the NSC's mass loss. Once it has formed, the NSC becomes comparable in mass to the halo’s enclosed mass at its radius. As a result, the NSC contributes to dynamical heating of the dark matter in the center of the halo. To properly model the impact of the NSC on the halo would require a more computationally involved self-consistent N-body treatment of the halo, but we examine some possible consequences in \S\ref{sec:fractions}.

Finally, we highlight that the relatively tight agreement among each of the realizations of the simulation can be attributed to the shared constraint on the total GC mass $M_{GC}$: in a realization involving a more massive progenitor at $t = 0$, fewer total GCs are present, and so the frequency of mergers is reduced. Correspondingly, realizations with less massive progenitors undergo many more mergers with individually less massive GCs, ultimately reaching similar final masses as the more massive progenitors. In this way, $M_{GC}$ regulates the properties of the NSC and final GC mass function.

\section{Discussion}
\label{sec:discussion}

After modeling UGC\,7369 over a wide range of galaxy and halo parameters, we have concluded that only models with NFW halos host enough GC mergers to create the observed NSC. We find further support for a cuspy profile and the merger-dominated formation pathway for the NSC in the final simulated mass function, as the mergers provide an explanation for UGC\,7369's lack of higher mass clusters compared to the universal distribution. For stellar masses $M_\star\sim 10^8\Msun$, a diverse variety of HI rotation curves have been observed (the ``diversity problem,'' e.g. \citealt{SantosSantos_Diversity} and references therein), including several profiles consistent with cusps. Thus, the detection of a cusp in UGC\,7369 is not in tension with observations of similar galaxies, although the sample of dwarf galaxies for which cusps or cores can be confirmed remains small. For this reason, the ability to determine the presence of a cusp or a core by studying GC populations is powerful, since GC populations may be characterized photometrically. Measuring galaxy rotation curves (for rotation-supported galaxies) and stellar velocity dispersions (for pressure-supported galaxies) requires spectroscopy, which is more difficult and observationally expensive.

With the goal of generalizing our analysis techniques beyond UGC\,7369 to complement kinematic-spectroscopic studies in other galaxies, in the following sections we discuss possible degeneracies in our models (\S\ref{sec:degeneracies}), list the requirements for concluding the presence of a cusp (\S\ref{sec:corecusp}), and motivate an extension of this analysis to a larger sample (\S\ref{sec:fractions}).

\subsection{Degeneracies and Robustness}
\label{sec:degeneracies}

We expect the initial masses, positions, and velocities of the GCs to constitute the main source of degeneracy with the halo features in terms of contributing to segregation and NSC formation effects. Because the friction timescale tends to increase with radius, if the GCs are initialized around a larger effective radius, the inspiral process is slowed, although some segregation effects should still be evident. As long as the typical friction timescale for the more massive GCs is no more than $\sim 10\,$Gyr at the effective radius used in the initialization process (see Figure \ref{fig:taudf}), though, the friction process should be able to take effect. We have confirmed that we find similar results for values of $R_e$ from $0.5 - 2\,$kpc. The effective (half-number) radius of GC systems is typically comparable to the $R_\mathrm{e}$ of the stellar component (see e.g. \citealt{Lim2018}. \citealt{Saifollahi2022}, and references therein), so we expect our mass segregation results to be fairly robust. The formation of an NSC should be similarly robust, although the final mass is slightly more sensitive to the initial $R_\mathrm{e}$ because the merger rate of the GCs is also slowed when they are initially spread over a larger volume. For example, the merger rates in the simulations presented in \cite{Chowdhury_GC_Nbody} and \cite{UDG1_DF} are significantly lower; an NSC never forms for the best-fit halo and GC population parameter choices because of the larger $R_\mathrm{e}$.

Besides variations in initial $R_\mathrm{e}$, an additional initial condition degeneracy to consider is that some mass segregation may have been present initially, e.g. due to variations in the cluster formation and/or accretion process. We expect evolution under dynamical friction to lead to further segregation, and so the segregation slope in the mass-projected radius plane alone may not be a robust predictor for halo properties. However, because NSC formation in cored halos is prevented primarily by core stalling, the presence or lack of an NSC and corresponding changes in the GC mass function at fixed total $M_{GC}$ should still distinguish between cuspy and cored halos as desired. Simulations including initial segregation are explored further in Appendix \ref{sec:appendixA}.

Finally, because the detectable differences between the simulated mass functions across different types of halos are primarily due to the effect of dynamical friction on GCs with $M \gtrsim 10^{5.5}\Msun$, our analysis is certainly dependent on the universal mass function's properties at higher masses: if all GCs are initialized with $M \lesssim 10^5\Msun$, fewer mergers will take place, and the NSC will not form. Proposed alternatives to the log-normal distribution include physically motivated power laws to describe the high mass tail (e.g. $n(M)\sim M^{-2}$ in \citealt{Elmegreen_GCMF}). While we have not carried out simulations involving other mass functions, because simulated draws from such distributions fitted to the high-mass tail of the observations tend to produce enough massive clusters to make up a similar $M_{NSC}$, we anticipate our conclusions will still hold.

\subsection{Implications for Studying the Core-Cusp Problem}
\label{sec:corecusp}

A cuspy halo is \textit{necessary} for the formation of an NSC in UGC\,7369, but looking ahead toward generalizations to other galaxies, we now consider additional criteria that, together with a cuspy halo, may be \textit{sufficient} for NSC formation. First, as discussed in \S\ref{sec:degeneracies}, a reasonably compact spatial distribution is required so that the massive GCs are able to inspiral within 10\,Gyr. We quantify this by requiring 
\begin{enumerate}
    \item A reasonably compact spatial distribution where $\taudf(M, R_\mathrm{e}, v_c(R_\mathrm{e})) \lesssim 10$\,Gyr
\end{enumerate}
for GCs with $M\sim 10^{5.5}\Msun$ which we expect to participate in the mergers. Note that $\taudf$ depends strongly on the halo parameters, so ``reasonably compact'' takes on different meanings at different halo masses. However, the similarity in friction timescales shown in Figure \ref{fig:taudf} for the varied halos we consider (summarized in Table \ref{tab:halos}) indicates that any galaxy falling on the typical mass-size relation with $M_\star\lesssim 10^{9}\Msun$ (e.g. Figure 9 of \citealt{ElvesI}) meets this criterion.

\begin{figure}
    \centering
    \includegraphics[width=0.48\textwidth]{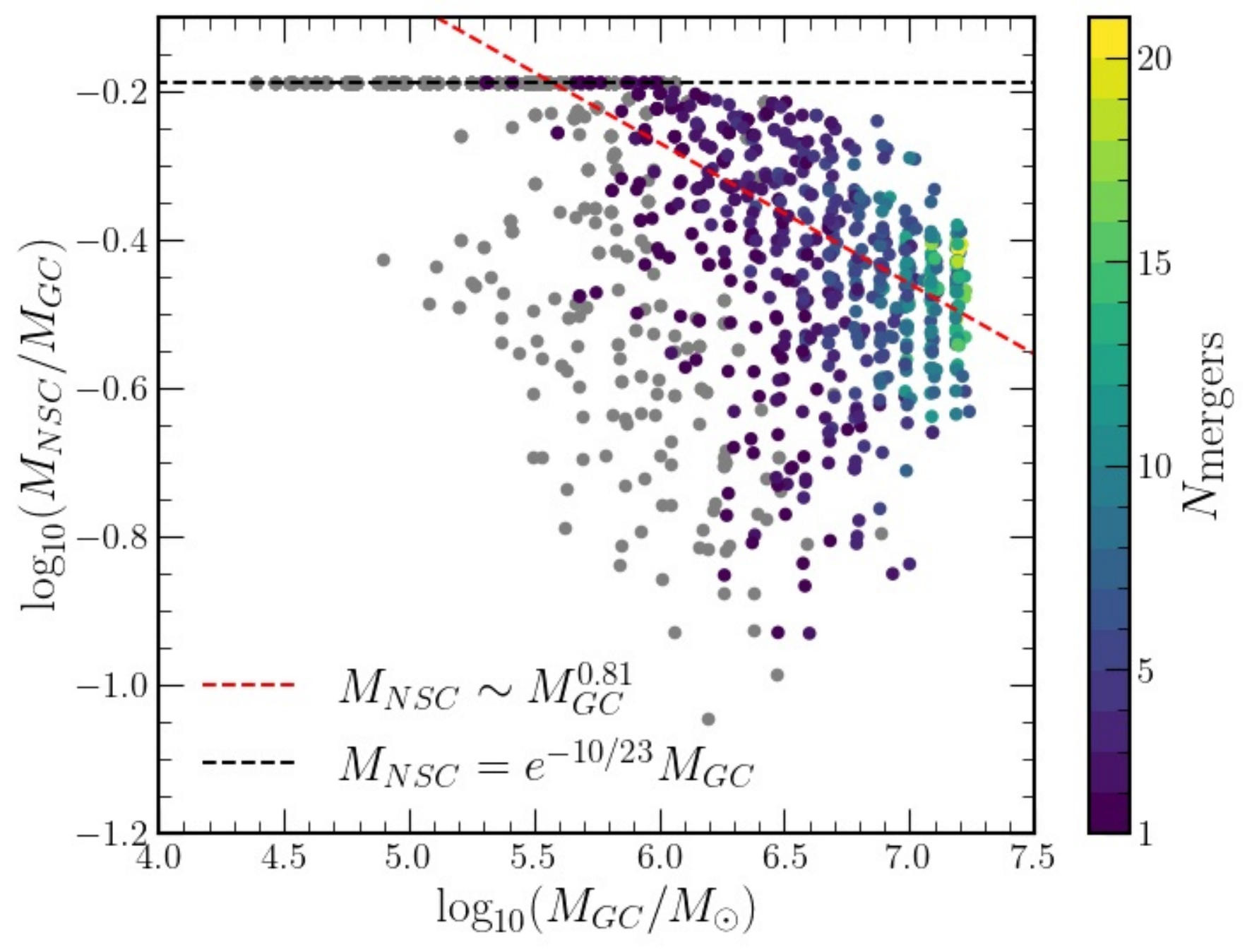}
    \caption{The mass of the NSC in the fiducial NFW halo as a function of varied initial total GC mass. The grey points are simulations that did not involve any mergers, and the remaining points are colored by number of mergers. The black dashed line indicates the expected NSC mass if all the initial GC mass ended up in the NSC (accounting for mass loss), and the red dashed line is a fit to all simulations in which at least one merger occurred.}
    \label{fig:NSC_vs_Mtot}
\end{figure}

Second, the formation of an NSC also depends on the total mass in GCs $M_{GC}$, since higher mass GCs are the progenitors of the NSC, and so enough of them must be present in the system. To test this, we conducted additional simulations in the fiducial NFW halo with varied total GC masses; Figure \ref{fig:NSC_vs_Mtot} demonstrates how the simulated NSC mass depends on the total $M_{GC}$.
At $10^4\Msun \lesssim M_{GC} \lesssim 10^{5.5}\Msun$, few GCs are present in the system, so the radial distributions are just stochastic, and few-to-zero mergers occur. From $10^{5.5}\Msun\lesssim M_{GC} \lesssim 10^6\Msun$, enough GCs are present that the radial distribution typically does contain information about the halo, but the NSC formation does not involve many mergers. Finally, above $M_{GC}\sim 10^6\Msun$, several mergers occur in the formation history of each NSC, and the trend with total GC mass correspondingly tightens considerably. Therefore, we conclude that another key requirement is
\begin{enumerate}
    \setcounter{enumi}{1}
    \item A total $M_{GC} \gtrsim 10^6\Msun$, so that multiple clusters massive enough to inspiral are likely to be present
\end{enumerate}

Ultimately, we propose the presence of an NSC as an important observable to identify cusps in galaxies meeting these two requirements, with stellar masses $M_\star\lesssim 10^9\Msun$, where observations indicate NSC formation is likely to be merger-dominated rather than in situ \citep{Neumayer_NSC}. Furthermore, because we do not expect GC mergers to occur frequently in cored profiles, we expect that the GC mass function in cored galaxies should match that of the ``universal'' form more closely than the distributions in cuspy galaxies, with the caveat that sufficiently spatially diffuse GC populations may prevent any differences from being apparent even in cuspy galaxies. 

\subsection{Extending to Larger Samples}
\label{sec:fractions}

More generally, at the population level, we therefore expect to find a correlation between the cuspy halo fraction $f_C$ and the nucleated fraction $f_N$, albeit to varying extents at different stellar masses, total GC masses, and environments. Unfortunately, it is challenging to explore this correlation with just the present \cite{Georgiev_sample1} sample: referring to Figure \ref{fig:MGC_Mstar}, we see that only $5$ galaxies meet our total GC mass requirement, out of which only three host GCs with masses $M > 10^{6}$, none of which are centrally located enough to be considered NSCs besides UGC\,7369. Within the sample, a majority of nucleated galaxies contain $< 5$ total GCs, making conclusions from analyses of mass functions and radial distributions harder to decouple from intrinsic stochastic variability.

However, a preliminary hint of this relationship is evident in several larger samples of dwarf galaxies. For the stellar mass range of interest, in low density environments, nucleation fractions are approximately $f_N\sim 0.4$ \citep{ElvesII}, consistent with estimates of cuspy fractions $f_C$ from rotation curve studies \citep{SantosSantos_Diversity}. Furthermore, in higher density environments, nucleation fractions tend to rise to $f_N\sim 0.8$ \citep{Neumayer_NSC, Hoyer2022, ElvesII}, and this trend of increasing nucleation fraction in cluster environments may be consistently explained. A commonly proposed solution to the core-cusp problem within $\Lambda$CDM invokes baryonic feedback: outflows generated by stars in the cusps may help transform the cusps into cores. Because observations suggest cluster dwarf galaxies are more rapidly and thoroughly quenched than their field counterparts (e.g. \citealt{Gunn_Gott, Masao_Burkert_2000, Grebel2003, Geha2012, Wetzel2015}), we would expect a higher cusp fraction $f_C$ in clusters, due to the reduced amount and efficiency of the feedback. More systematic studies of GC systems in dwarf galaxies in such environments are needed to verify this possible $f_C$--$f_N$ correlation.

Although this picture coherently explains many observations, we also note some prominent exceptions to the NSC formation requirements listed in \S\ref{sec:corecusp} that should be carefully accounted for when curating an extended sample. With regard to the first requirement on total GC mass, some dwarf galaxies have been observed to host NSCs with $M_{NSC} < M_{GC} \lesssim 10^6\Msun$ (\citealt{NSC_formation_dwarfs}). Although these NSCs tend to have a high fraction of old, metal-poor stars consistent with a merger dominated formation history, they may also be attributed to either friction-induced inspiral of just one massive GC as allowed by the tail of the mass function, or a smaller number of mergers of lower-mass GCs that were by chance initialized much closer to the halo's center. It is much more difficult to extract information about the shape of the halo in these cases due to the small number of GCs.

With regard to the second requirement on spatial extent, some ultra-diffuse galaxies (UDGs) are known to be nucleated \citep{Virgo_UDGs}, despite having a large $R_\mathrm{e}$. If we assume that the same friction and merger processes govern NSC formation in these systems, then we can either (i) place strong upper bounds on total halo mass, so that the friction timescale is within 10\,Gyr at the larger radius, (ii) constrain the initial radius of the GCs to be smaller, perhaps at an earlier time in the galaxy's history before its stellar population became so extended, or (iii) claim that the initial mass function of the GCs is more top-heavy than the standard ``universal'' form. Either way, these objects certainly merit further study.

Finally, we emphasize that in some especially-GC-rich galaxies, capable of producing a massive NSC, this correlation holds only \textit{while the NSC forms}, and not afterward. The reason for this is that if the mass of the NSC becomes comparable to the mass of the halo within its orbital radius, it may contribute significantly to the evolution of the halo through dynamical heating. By exchanging energy with the dark matter in the center of the halo, the NSC can flatten the cusp by spurring dark matter to larger orbital radii \citep{ElZant2001, Elzant2004}. Such NSC-induced heating may be an important dynamical solution to the core-cusp problem \citep{Goerdt2010, ArcaSedda_Lack_of_NSCs_BHs}, and could also explain observations of NSCs in galaxies otherwise well-fit by cores: it is required that the NSC form in a cusp, but it may subsequently produce a core. In UGC\,7369, for example, because the individual GCs do not dominate over the halo (as measured by the ratio of each GC’s mass to the enclosed halo mass at its radius) prior to the NSC’s formation, a static cuspy or cored halo is a reasonable approximation. However, after the NSC has formed (in approximately $4\,$Gyr; see Figure \ref{fig:NSC_evolution}), its mass does exceed the enclosed halo mass at its radius. Accurately capturing the possible subsequent formation of a core requires self-consistent N-body treatments of the halo, which we defer to a later work.

\section{Summary}
\label{sec:summary}

We have demonstrated that dynamical friction can naturally explain the observed mass segregation in UGC\,7369, and furthermore, that it can also account for the presence of an NSC and corresponding difference between the galaxy's GC mass function and the universal form, as long as an NFW halo density profile is assumed. Because no NSC can form in cored halos over a wide range of parameters, we take this as strong evidence that UGC\,7369's halo is cuspy.

The NSCs in our simulations form via mergers that take place near the centers of the halos, after being driven to the center through dynamical friction processes, which act very rapidly within the first $\sim 3$\,Gyr on the most massive clusters. In a cored halo, their progress toward the center is halted by core stalling, and so no mergers occur, which allows the presence of an NSC to be an important indicator in constraining the halo shape. More generally, in this merger-dominated formation regime, the shape of the GC mass function can be thought of as a new observable, as it is altered by the depletion of the higher mass clusters that are more susceptible to infall and merging.

Our results regarding NSC formation are robust against significant variations in halo parameters and initial conditions, including adjustments to the initial radius at which GCs tend to form, reasonable degrees of initial mass segregation, and halo parameters. The most important observational input to the model is the total mass in GCs, which regulates the number of clusters through our assumed log-normal universal GC mass function. Our results should also remain robust to a variety of changes in the mass function, though they are reliant on a sufficient number of clusters being initialized with a high mass, as only the high-mass clusters have short enough friction timescales.

Potential improvements to the modeling could include using a ``live'' halo with a large number of massive particles so that no semianalytic prescription for dynamical friction would be necessary. Additionally, treating the GCs themselves as live N-body systems would provide a better treatment of the internal dynamics, which will be especially important following mergers. Outside of these more computationally expensive improvements, the most beneficial update to our simulations would involve improving the semi-analytic dynamical friction implementation. Although our current calculations for the Coulomb logarithm and friction cutoffs are physically motivated and have been shown to match N-body results, the Chandrasekhar formula is still an intrinsically rough approximation, and as core stalling is the primary driving phenomenon behind our results, we could certainly make more precise constraints with a better prescription.

In order to extend our results, the most important observations would involve GC-rich galaxies with stellar masses spanning the range $10^7\Msun \lesssim M_\star \lesssim 10^9\Msun$ at which we expect NSCs to be formed through mergers, both with and without nucleation. Besides new observations, analyses of archival data that further reduce contamination and provide more complete coverage of GCs would allow for this modeling technique to be applied to additional galaxies and provide information about halo shape at a population level. Further analysis of these population statistics is the clearest next step for using GC populations to study the core-cusp problem: with a larger sample, the relationship between the cuspy and nucleated fraction would become more evident, and it should also be possible to discern possible differences between the GC mass functions of nucleated and non-nucleated galaxies.

\section*{Acknowledgments}
We are very grateful to Nitsan Bar for many insightful discussions, and to Jiaxuan Li for carrying out the S\'ersic fitting. We also thank Scott Tremaine for his helpful comments, and the anonymous referee for valuable input. S.M. acknowledges support from the National Science Foundation Graduate Research Fellowship under Grant No. DGE-2039656. S.D. is supported by NASA through Hubble Fellowship grant HST-HF2-51454.001-A awarded by the Space Telescope Science Institute, which is operated by the Association of Universities for Research in Astronomy, Incorporated, under NASA contract NAS5-26555. J.E.G. gratefully acknowledges support from NSF grant AST-1007052.

\software{REBOUND (\citealt{rebound}, \citealt{reboundias15}, \citealt{reboundx}), numpy \citep{numpy}, matplotlib \citep{matplotlib}, imfit \citep{imfit}}.

\appendix
\section{Simulations with Initial Mass Segregation} \label{sec:appendixA}

\begin{figure*}
    \centering
    \includegraphics[width=0.98\textwidth]{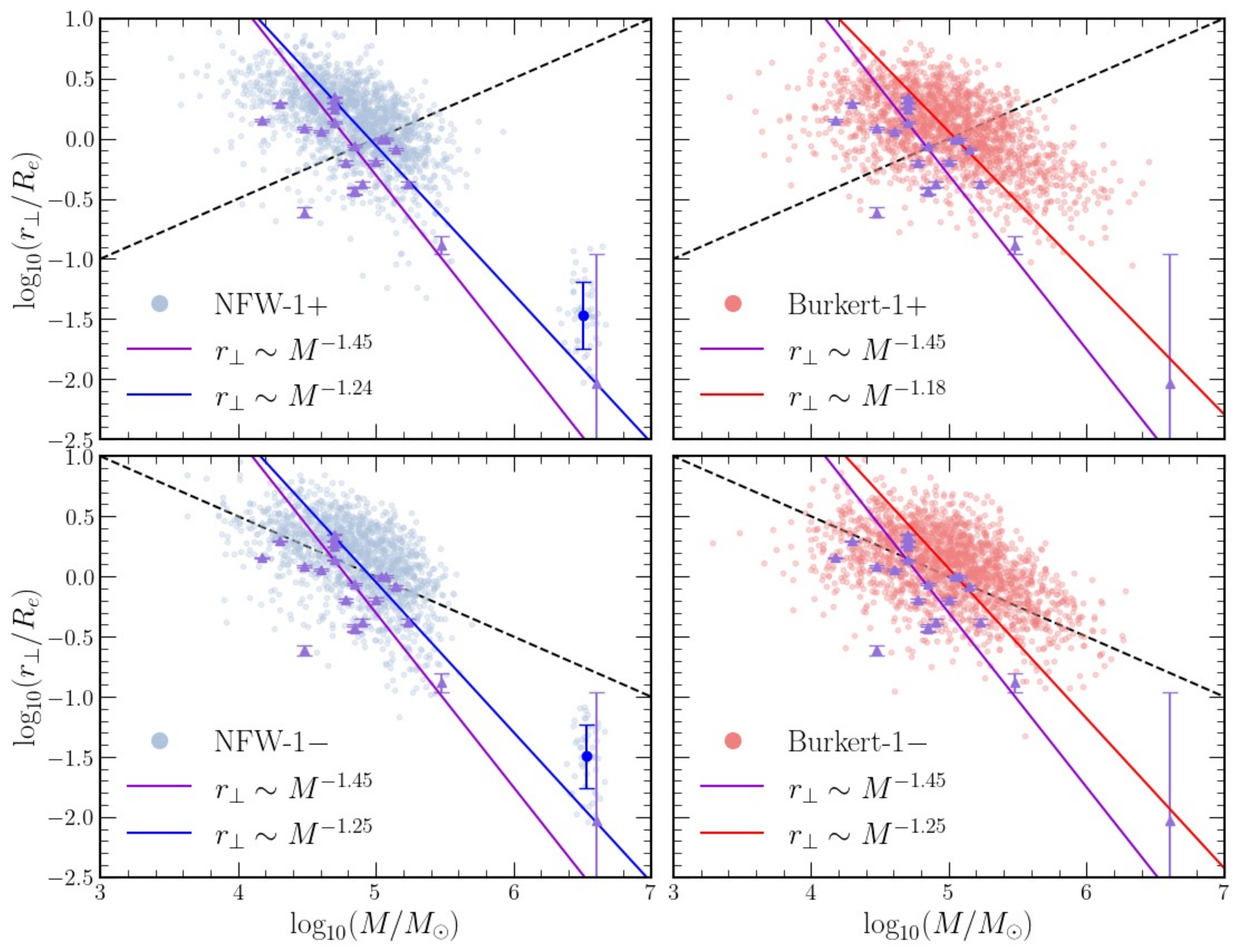}
    \caption{Similar to Figure \ref{fig:rperp_vs_M}, but for the models with initial mass segregation, with the average initial radius as a function of mass shown as the black dashed line. Again, each of the lighter colored blue or red points respectively is one GC after 10\,Gyr of evolution, and points from all 50 realizations are shown, in addition to the observed GCs in purple. Linear regressions to the points demonstrating the segregation are also included, and the dark blue point on the NFW panels shows the mean properties of the NSC. Mass errors on the observed points and projection errors on the simulated points are suppressed for clarity.}
    \label{fig:initial_segregation}
\end{figure*}

To study how initial mass segregation (e.g. due to GC formation efficiency varying at different radii) could impact our conclusions, we carried out simulations in the fiducial NFW and Burkert halos with an adjusted initialization procedure: for each GC, after drawing an initial radius $r_0$ and velocity $v_0$ from the density profile and distribution function as usual, we added a random logarithmic radial shift drawn from a normal distribution,
\begin{equation}
    \delta (\log_{10}(r)) \sim \mathcal{N}(\mu = \alpha\log_{10}(M/10^5\Msun),\; \sigma^2=0.01)
\end{equation}
so that $\log_{10}(r_i) = \log_{10}(r_0) + \delta(\log_{10}(r))$. We chose $\alpha = \pm 0.5$ for either positive or negative initial segregation slopes respectively, to mimic the approximate degree of segregation that occurred in the rest of our simulation suite after 10\,Gyr of evolution. Next, we modify the initial speed, 
\begin{equation}
    v_i = \sqrt{v_0^2 + 2(\phi_h(r_0)-\phi_h(r_i))}
\end{equation}
so that the total initial energy is conserved. The GC is then initialized with radius $r_i$ and speed $v_i$, and angles for both position and velocity drawn isotropically. Figure \ref{fig:initial_segregation} shows the simulation initial conditions and evolution after 10\,Gyr in the $(\log_{10}(r_\perp/R_\mathrm{e}), \log_{10}(M/\Msun))$ plane, analogous to Figure \ref{fig:rperp_vs_M}, and Table \ref{tab:initial_segregation} summarizes the results.

\begin{table*}
    \centering
    \begin{tabular}{c|c|c|c|c|c|c}
        ID & Mass Segregation Slope & $\langle N_{GC}\rangle$ & $\langle N_{\textrm{mergers}} \rangle$ & $\langle M_{NSC} \rangle$ $(10^5\Msun)$ & $\langle r_{\perp\textrm{, NSC}}/R_\mathrm{e} \rangle$ & Notes \\
        \hline
        Observations & $-1.4\pm0.3$ & 21 & N/A & $40\pm16$ & $0.010\pm0.010$ & N/A\\
        NFW-1 & $-1.24\pm0.03$ & $31\pm7$ & $7\pm3$ & $32\pm7$ & $0.04\pm0.02$ & fiducial \\
        NFW-1$+$ & $-1.24\pm0.03$ & $32\pm7$ & $7\pm2$ & $33\pm6$ & $0.04\pm0.03$ & $\alpha=+0.5$ \\
        NFW-1$-$ & $-1.25\pm0.03$ & $30\pm5$ & $8\pm2$ & $34\pm5$ & $0.03\pm0.018$ & $\alpha=-0.5$ \\
        Burkert-1 & $-1.22\pm0.04$ & $38\pm9$ & $0.10\pm0.3$ & N/A & N/A & matching NFW-1 \\
        Burkert-1$+$ & $-1.18\pm0.04$ & $37\pm8$ & $0.1\pm0.5$ & N/A & N/A & $\alpha=+0.5$ \\
        Burkert-1$-$ & $-1.25\pm0.04$ & $38\pm7$ & $0.2\pm0.4$ & N/A & N/A & $\alpha=-0.5$\\ \hline
    \end{tabular}
    \caption{A summary of simulation results with initial mass segregation with the same columns as Table \ref{tab:results_summary}, and the observation, fiducial NFW-1, and Burkert-1 halo results copied for easier comparison.}
    \label{tab:initial_segregation}
\end{table*}

We find that initializing the GCs with some mass segregation in this manner does not significantly change our conclusions. While we do find that introducing initially negative mass segregation slightly increases the number of mergers in both NFW and Burkert halos, and correspondingly slightly increases the mass of the resulting NSC in cuspy halos, all initially segregated results are consistent with their respective fiducial cases. Because the friction timescales for clusters with $M \gtrsim 10^{5.5}\Msun$ are well within 10\,Gyr, there is plenty of time for the NSC to form even with initially positive segregation, and lower mass clusters are still not as strongly impacted despite starting at closer-in radii.

\bibliographystyle{aasjournal}

\end{document}